\documentclass[journal]{IEEEtran}
\usepackage{multirow}
\usepackage{color}
\usepackage{amsmath}
\usepackage{algorithm}
\usepackage{algpseudocode}
\usepackage{newtxtext,newtxmath,amsmath}
\usepackage{subcaption}
\usepackage{graphicx}
\usepackage{balance}
\usepackage{textcomp}

\newcommand\textblue[1]{\textcolor{blue}{\textbf{#1}}}
\newcommand\textred[1]{\textcolor{red}{\textbf{#1}}}

\begin{document}

\title{Universal Stego Post-processing for Enhancing Image Steganography}

\author{Bolin Chen, ~ 
              Weiqi Luo*, ~\IEEEmembership{Senior Member, ~IEEE, }
              Peijia Zheng, ~\IEEEmembership{Member, ~IEEE, }
              Jiwu Huang,~\IEEEmembership{Fellow,~IEEE}
        
\thanks{This work was presented in part at the 7th ACM Workshop on Information Hiding and Multimedia Security, Paris, France, July 2019.}
        
\thanks{B. Chen, W. Luo and P. Zheng are with the School of Data and Computer Science, and Guangdong Key Laboratory of Information Security Technology, Sun Yat-sen University, Guangzhou 510006, P.R. China. Corresponding author: Weiqi Luo (e-mail: luoweiqi@mail.sysu.edu.cn).}

\thanks{Jiwu Huang is with the College of Information Engineering, Shenzhen University, Shenzhen 518052, P.R. China.}
}

\markboth{Journal of \LaTeX\ Class Files,~Vol.~14, No.~8, August~2015}%
{Shell \MakeLowercase{\textit{et al.}}: Bare Demo of IEEEtran.cls for IEEE Journals}

\maketitle
 
\begin{abstract}

It is well known that the designing or improving embedding cost  becomes a key issue for current steganographic methods.  Unlike existing works,  we propose a novel framework to enhance the steganography security via  post-processing on the embedding units (i.e., pixel values and DCT coefficients) of stego directly.   
In this paper,  we firstly analyze the characteristics of STCs (Syndrome-Trellis Codes),  and then design the rule for post-processing to ensure the correct extraction of hidden message.  Since the steganography artifacts are typically reflected on image residuals,  we try to  reduce the residual distance between cover and the modified stego in order to enhance steganography security.  To this end, we model the post-processing as a non-linear integer programming, and  implement it via  heuristic search.   In addition,  we carefully determine  several important issues in  the proposed post-processing, such as the candidate embedding units to be modified,  the direction and amplitude of post-modification, the adaptive filters for getting residuals, and the distance measure of  residuals.  Extensive experimental results evaluated on both hand-crafted steganalytic features and deep learning based ones demonstrate that the proposed method can effectively enhance the security  of most modern  steganographic methods both in spatial and JPEG domains.  
\end{abstract}


\IEEEpeerreviewmaketitle

\section{Introduction}

\IEEEPARstart{I}{mage} steganography is a technique to hide secret message into cover images via modifying some image components in an imperceptible manner. On the contrary, image steganalysis aims to detect the existence of secret message hidden by image steganography.  During the past decade,  many effective steganography methods have been proposed with the development of the steganalytic techniques. 

Image steganography can be divided into two categories, that is, spatial steganography and JPEG steganography.  In modern research,  both of them are usually designed under the framework of distortion minimization \cite{fridrich2007practical}, in which the design of  embedding cost is the key issue.  Typically,  the embedding cost tries to measure the statistical detectability of each embedding unit (i.e. pixel or DCT coefficient). The smaller the embedding cost, the more likely the corresponding unit will be modified during the subsequent  operation of Syndrome-Trellis Codes (STCs)  \cite{filler2011minimizing}.  Up to now, there are many effective cost  have been proposed in spatial domain. Most of them such as HUGO \cite{pevny2010using}, WOW \cite{holub2012designing}, S-UNIWARD \cite{holub2014universal}, HILL \cite{li2014new} and MIPOD \cite{sedighi2016content} adopt an additive cost, meaning that they assume the embedding impact for each unit is independent. Some methods such as CMD (Clustering Modification Directions) \cite{li2015strategy}, Synch \cite{denemark2015improving} and DeJoin \cite{zhang2017decomposing} improve the existing  additive cost  via sequentially embedding message and  updating the cost to synchronize the modification direction.  These methods usually achieve better security performance since  the mutual impacts of adjacent embedding units are taken into consideration.   
For  JPEG steganography, the additive cost-based  methods include UED \cite{guo2014uniform}, J-UNIWARD \cite{holub2014universal}, UERD \cite{guo2015using}, BET \cite{hu2018efficient}, and the  non-additive one includes  BBC \cite{li2018defining}, which aims to preserve the spatial continuity at block boundaries.  To enhance security, some other steganography methods aim to adjust existing costs via  highlighting the details in an image \cite{chen2016defining, chen2018defining} or reassigning lower costs to controversial units \cite{zhou2017new, zhou2018controversial}.  Recently, some deep learning techniques such as Generative Adversarial Network (GAN) \cite{goodfellow2014generative} and adversarial example \cite{szegedy2013intriguing} have been applied in  steganography. For instance, ASDL-GAN \cite{tang2017automatic} and UT-GAN \cite{yang2019embedding, yang2019towards} can learn costs that are directly related to the undetectability against the steganalyzer.  ADV-EMB \cite{tang2019cnn} and  method \cite{bernard2019exploiting} adjust the costs according to the gradients back-propagated from the target Convolutional Neural Network (CNN)-based steganalyzer.

Note that  above steganography methods mainly focus on designing embedding costs,  and usually  employ the STCs to minimize the total costs in subsequent data hiding.  However,  most existing embedding costs seem  empirical,   which would not be effective to  measure the statistical detectability of  embedding units.  In addition,  minimizing  the total costs using STCs  would not always produce high security stegos. 
Unlike existing works,  we  propose  a novel framework to enhance the security of current steganography methods both in spatial and JPEG domains via stego post-processing, which aims to reduce the residual distance  between  cover and modified stego.   We firstly formulate the stego post-processing as a non-linear integer programming problem,  and solve it using a heuristic search method - Hill Climbing.   To achieve good security performance,  the adaptive filters for obtaining image residuals and the distance measure  are carefully designed.   In addition,  four acceleration strategies according to the characteristics of post-modification  are considered  to speed up our algorithm.   Experimental results show that the proposed method  can significantly enhance the security performance of  the existing steganography methods, especially when the payloads and/or quality factors are large. Note that this paper is an extension of our previous work \cite{chen2019enhancing}. Compared to our preliminary work \cite{chen2019enhancing}, the main differences of this paper are as follows: 1)  Instead of using a fixed filter in  \cite{chen2019enhancing} to obtain image residuals, in this paper, we carefully design multiple adaptive filters which can better suppress image content while preserve the artifacts left by data hiding;   2)  The method in this extended version significantly accelerates the post-processing via restricting the position of modified units,  the direction and amplitude of modification,  and adopting a fast method for convolution; 3)  More extensive experimental results and analysis are given in this paper.  For instance,   both conventional and deep learning  steganalytic models  are used for security evaluation.  In addition to BOSSBase \cite{bas2011break},  other two public databases (i.e., BOWS2 \cite{bows2} and ALASKA \cite{cogranne2019the}) which include 90,005 images are used for evaluation.   We provide more analysis on statistical characteristics of post-modification and the processing time.  In addition,  both spatial and JPEG steganographic methods are considered in this paper;  4) The extensive experimental results show that the proposed method can achieve higher security than the work \cite{chen2019enhancing}.

Compared to those works (e.g. CMD \cite{li2015strategy},  BBC \cite{li2018defining},  and methods \cite{chen2016defining, chen2018defining, zhou2017new, zhou2018controversial}) which  also aim to enhance existing steganography methods,   the main differences of the proposed method are as follows:
\begin{itemize}
\vspace{0.5em}
\item  First of all,  almost all  related works  try to  modify embedding costs of existing steganography during data hiding,  while the proposed method tries to modify embedding units (pixel values or DCT coefficients) directly after the data hiding with the existing steganopgraphy is completed.  Note that the early steganography  OutGuess \cite{provos2001defending}  divides the cover into two non-overlapping parts: one part for data hiding, the other part for histogram correction.  From this point of view,  OutGuess can be regarded as an enhanced steganography based on post-processing.  However,  the  embedding capacity of OutGuess is significantly reduced since it has to reserve a relatively high  proportion of embedding units for histogram correction. What is more, it can be easily detected by the modern steganalytic methods based on higher order statistics.  

\vspace{0.5em}
\item The principle of related works is quite different.  For instance,  CMD  aims at clustering modification directions and BBC aims at preserving the block boundary continuity, OutGuess \cite{provos2001defending}  aims at preserving the histogram, while the proposed method aims at reducing the residual distance between cover and the resulting stego.  Since the proposed method is performed on the stego,  it is not contradictory to those  steganographic works based on STC embedding.  As shown in  Section \ref{sec:experiments},  most modern steganographic methods, such as MIPOD, CMD-HILL, J-UNIWARD and  BET-HILL,  can be further improved after using the proposed post-processing.

\vspace{0.5em}
\item Most existing works are usually designed under a special domain.  For instance,  the spatial steganographic methods such as MIPOD and CMD are difficult to be adopted in JPEG with satisfactory results. Similarly, those effective works for JPEG steganography may not effective in spatial domain. Comparatively,  the proposed method is effective in both domains. 
\end{itemize}

The rest of this paper is arranged as follows. Section \ref{sec:preliminaries} describes  STCs and their robustness against post-modification.  Section \ref{sec:method} describes the proposed framework.  Section \ref{sec:experiments} presents experimental results and discussions.  Finally, the concluding remarks  of this paper are given in Section \ref{sec:conclusion}.

\section{Robustness Analysis of STCs}
\label{sec:preliminaries}

Most  current steganographic methods are constructed under the framework of distortion minimization.  After the embedding costs are carefully designed,  some coding methods are then used to embed secret message into  cover image in order to minimize the total cost.   In practical applications,  STCs  is  widely used  in modern steganography methods both in spatial and JPEG domains.  Since the  extraction of hidden message after using the proposed method is related to STCs,  we will give a brief overview of STCs and its robustness against post-modification in the following. 

\subsection{Review of Syndrome-Trellis Codes}
STC is one of the popular coding methods which is able to embed secret message into the cover image efficiently while approaching the optimal coding performance. It can be used to solve binary or non-binary embedding problem under the steganography framework of distortion minimization. For binary problem, the message embedding and extraction for spatial steganography \footnote{Similar results can be obtained for JPEG steganography.} can be formulated as follows:

\begin{equation}
\label{eq:embed}
Emb(X, m) = \arg \min_{ P(Y) \in C(m) } D(X, Y)
\end{equation}
\begin{equation}
\label{eq:extract} 
Ext(Y) = \mathbb{H}  P(Y)
\end{equation}

where $Emb()$ is the function for data embedding. $Ext()$ is the function for message extraction. $ X $ is a cover image. $ Y $ is a stego image. $ m $ is a secret message. $ P $ is a parity function such as $ P(Y) = Y \mod 2 $. $ \mathbb{H} $ is a parity-check matrix of a binary linear code $ C $. $ C(m) = \{ z | \mathbb{H} z = m \} $ is the coset corresponding to syndrome $ m $. STCs constructs the parity-check matrix $ \mathbb{H} $ by placing a small submatrix  $ \hat{ \mathbb{H} } $ along the main diagonal. The height of the submatrix $ \hat{ \mathbb{H} } $ is a parameter that can be used to balance the algorithm performance and speed. Using parity-check matrix $ \mathbb{H} $ constructed in this way, equation (\ref{eq:embed}) can be solved optimally by Viterbi algorithm with linear time and space complexity w.r.t. $ n $, which is the dimension of $ X $.

For the q-ary (q $>$ 2) embedding problem, STCs solves it efficiently via multi-layered construction. It decomposes the q-ary problem into a sequence of similar binary problem and then applies the above solution for binary problem. The q-ary problem can be solved optimally if each binary problem is solve optimally.  Refer to  \cite{filler2011minimizing} for more details of STC.

\subsection{Analysis of Robustness of STCs}
\label{subsec:robustness}
From equation (\ref{eq:extract}), the value of extracted message is determined by $ \mathbb{H} $ and $P(y)$. In a covert communication, since $ \mathbb{H} $ is fixed for a cover image, the message extraction completely relies on $P(y)$.
Therefore, if there exists a modification matrix $\Delta$ such that $P(Y+\Delta) = P(Y)$,  we can extract exactly the same secret message from $ Y + \Delta $ and $ Y $, which shows the robustness of STCs against the modification $\Delta$ in this case.  Generally, image steganography embeds message into lower bits of the cover image for not introducing visually perceptible artifacts. Therefore, the parity function  $P$ of q-ary STCs returns the $ 1^{st} $ to $ k^{th} $ LSBs of the input image, where $ k = \lceil \log _2 q \rceil $. Based on this characteristic, q-ary STCs' robustness against post-modification can be formulated as follows:

\begin{equation}
\label{eq:robustness}
Ext(Y) = Ext(Y + \Delta), \quad \Delta_{ij} = 2^k \times n_{ij}, n_{ij} \in \mathcal{Z}
\end{equation}
where $Y$ and $ \Delta $ are  matrices of the same size  $ n_1 \times n_2 $, $ \Delta_{ij}$ denotes the $ ij^{th}$ element of the modification matrix $\Delta$

Taking a stego image $Y$ obtained by  ternary STCs (i.e. $ q = 3 $) for example, in this case, $ k = \lceil \log _2 q \rceil = 2 $.   $ \Delta_{i,j} = 2^2 \times n_{ij} = 4n_{ij} $, $ n_{ij} \in \mathcal{Z} $. Therefore,  we conclude that  adding a multiple of 4 to any elements of the stego image will not confuse the message extraction at all.

\section{Proposed Framework and Method}
\label{sec:method}

In this section, we first describe the  framework of stego post-processing, and then present some implementation details, including the selection of some important parameters and four strategies to speed up processing.  Finally,  we will give the full description of the proposed algorithm under this framework. 

\begin{figure}
\centering
\includegraphics[width=\linewidth]{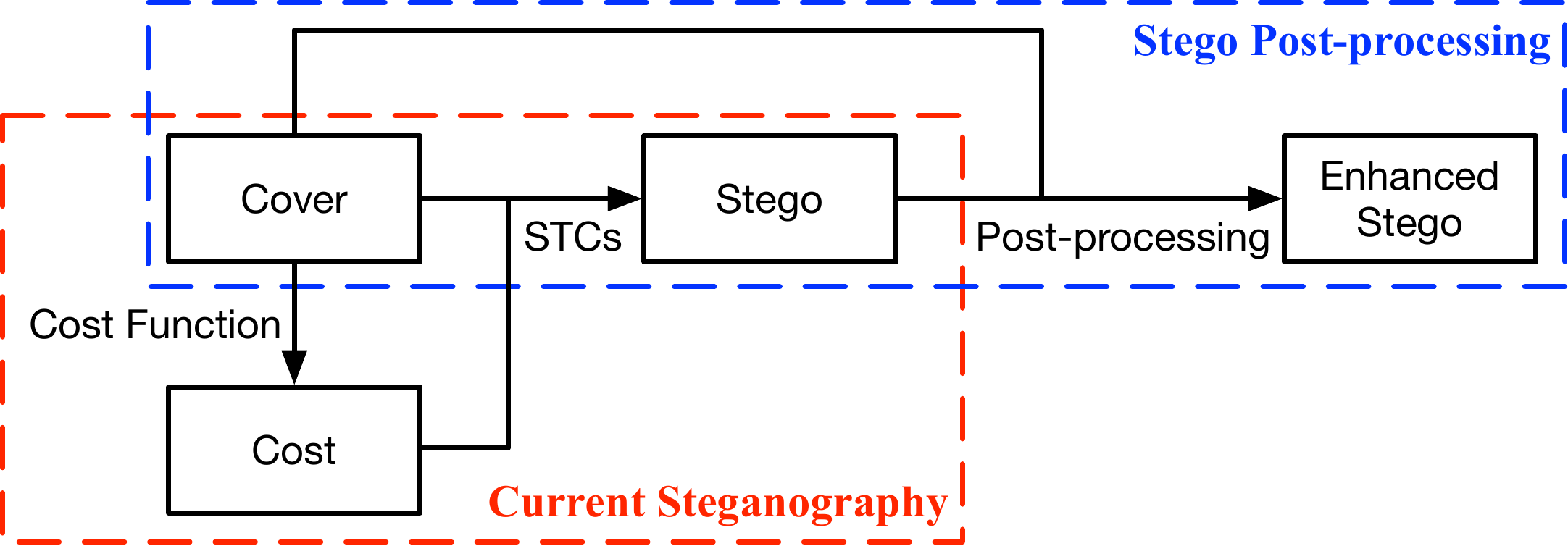}
\caption{The framework of modern steganography  vs.  the proposed stego post-processing.}
\label{fig:postprocess_in_stego}
\end{figure}

\begin{figure}[t!]
\centering
\includegraphics[width=\linewidth]{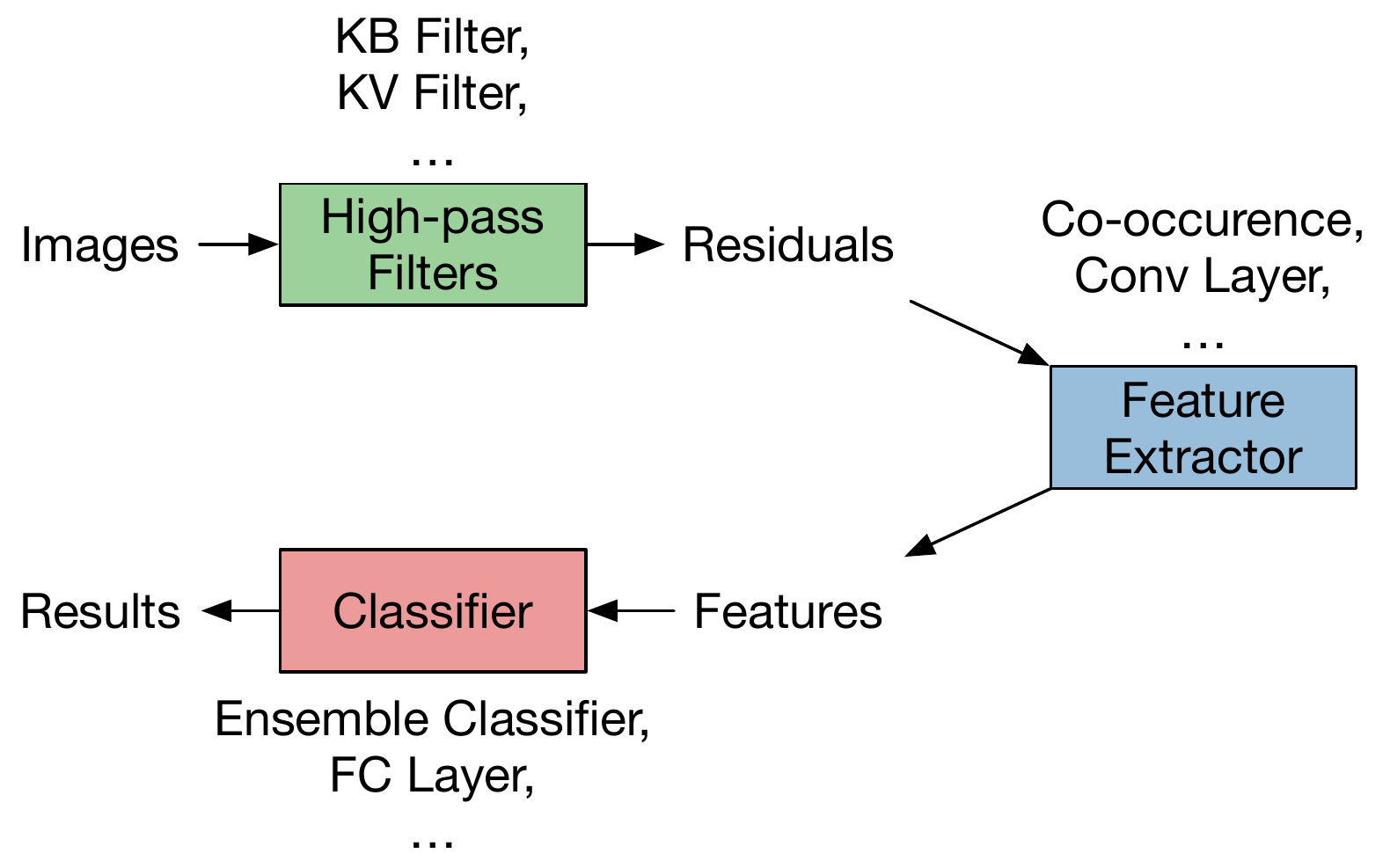}
\caption{The framework of modern image steganalytic methods.}
\label{fig:framework_steganalysis}
\end{figure}

\subsection{Framework of Stego Post-processing}
As shown in Fig. \ref{fig:postprocess_in_stego}, the current steganography firstly designs costs for all embedding units of a cover image, and then uses STCs to embed secret message into cover to get the resulting stego.  Quite different from  the existing works,  the proposed framework aims to enhance the steganography security via reducing image residual distance between cover and stego using stego post-processing.   Since most current steganography methods, such as  HILL \cite{li2014new} and UERD \cite{guo2015using}, employ ternary STCs for data embedding,   the ternary case (i.e. $k=2$) is considered in our experiments.  Please note that it is easy to extend our method for different  $k$. 

\subsubsection{\textbf{Main Idea of Stego Post-processing}}
\label{subsub:Main_Idea}

It is well known  that the steganography will introduce  detectable artifacts into  image residuals, and thus most  effective steganalyzers based on hand-crafted features (e.g. SRM \cite{fridrich2012rich}, GFR \cite{song2015steganalysis}) and deep learning (e.g. Xu-Net \cite{xu2016structural}, Ye-Net \cite{ye2017deep}, and J-Xu-Net \cite{xu2017deep}) are mainly based on analyzing  image residuals in spatial domain.  As illustrated in Fig. \ref{fig:framework_steganalysis},  these steganalytic methods usually contain 3 components, that is,  high-pass filters to obtain image residuals, feature extraction operator of image residuals and a classifier based on the features.  Since the steganography signal is rather weak compared to  image content,  good high-pass filters can effectively suppress image content and improve the signal-to-noise ratio (note that for steganalysis, noise here is image content),  which is very helpful for steganalysis.  From this point of view,  if the image residual distance between cover and modified stego image is smaller,  the security performance is expected to be better.  Therefore,  the main idea of the proposed framework is to reduce such distance via stego post-processing.  Combined with the robustness analysis on STCs in section \ref{subsec:robustness},  the proposed stego post-processing can be formulated as the following optimization problem: 

\begin{equation}
\begin{aligned}
& \underset{Z}{\text{minimize}}
& & Dist(Res(Z), Res(X)) \\
& \text{subject to}
& & Z = Y + 2^2 \times N, \\
&&&  N \in \mathcal{Z}, \\
&&&  Z \in \mathcal{V}.
\end{aligned}
\end{equation}

where  $ Res(X) $ \footnote{For spatial steganography, $X, Y$ and $Z$ denote pixel values of the corresponding images.   For JPEG steganography,  they denote the DCT coefficients.  The image residual is obtained and analyzed in spatial domain both for spatial and JPEG steganography.  } denotes the image residual of image $X$ in spatial domain,  $ Dist(Res(Z),Res(X)) $ denotes the distance between two image residuals $Res(Z)$ and $Res(X)$;   $ X $ is a cover image,  $ Y $ is a stego image obtained with an existing steganography method, $ Z $ is a modified version of $Y$ with our post stego-processing;  $N$ is an integer matrix;   $ \mathcal{V} $ denotes the available range of embedding units of $Z$. Taking spatial steganography for instance,  every unit in  $ \mathcal{V} $ should be an integer  in the range of $[0, 255]$.

Note that the proposed  framework tries to modify a resulting stego $Y$  obtained with an existing steganography method under the framework of distortion minimization,  thus any modification on $Y$ will inevitably increase the total distortion.  However,  we  expect that the steganography security  would  become better since the residual distance between cover $X$ and the resulting stego  $Z$ is reduced after stego post-processing.

\begin{figure}[!]
\centering
\includegraphics[width=0.85\linewidth]{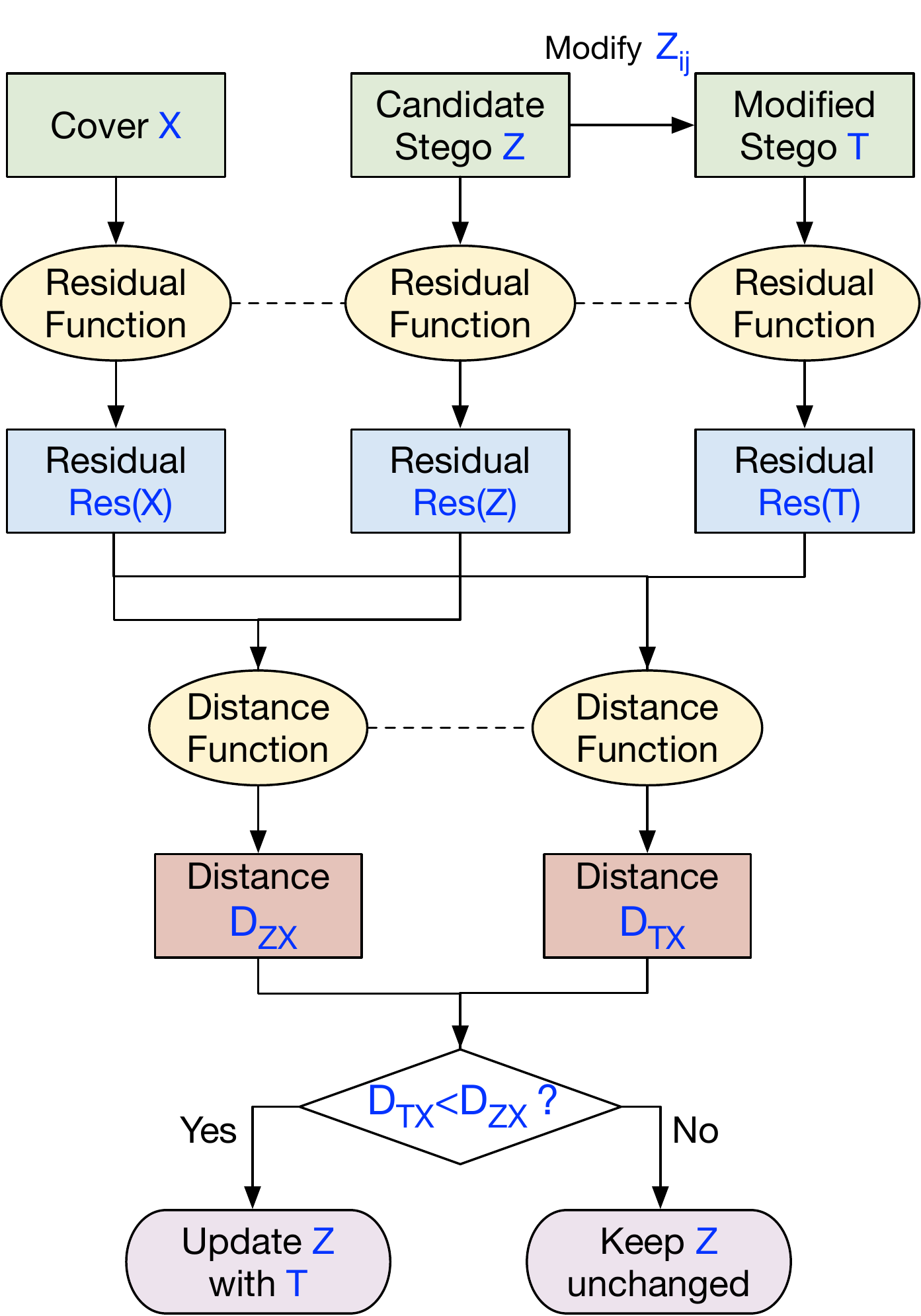}
\caption{The proposed method to  update a target embedding unit  within a stego image.}
\label{fig:spp_framework}
\end{figure}

\vspace{0.5em}\subsubsection{\textbf{Implementation of the Framework}}

Since the modifications are limited on integers, and the distance function $ Dist( ) $ is usually non-linear,  the optimization problem  described in  the previous section is a non-linear integer programming, which is very hard to find the optimal solution.  In our experiments,  we employ a  greedy algorithm, i.e.,  Hill Climbing,  to find an approximate solution.  Specifically,  from an initial stego $Y$,  we sequentially process the  embedding units  one by one  to  iteratively  reduce its residual distance to  cover $X$   until all embedding units are dealt with. 

Fig. \ref{fig:spp_framework} illustrates how the proposed method  updates a target embedding unit  within a stego image. 
Let $X$ denote cover,  $Z$ denote the candidate stego which is initialized as the stego $Y$  with an existing steganography method,  $T$ denote the temporary variable for the modified version of $Z$ after changing a target unit $Z_{ij}$  according to the rule  described in section \ref{sec:preliminaries}  (i.e. $T=Z, T_{ij}=T_{ij}+2^2\times n,n\in \mathcal{Z}$).  By doing so, we can assure that  the secret messages extracted from $Z$  and $Y$ are exactly the same after modification.  
To determine whether the modified stego $T$ is better than the candidate one $Z$,  we firstly  apply $ Res() $ function on cover image $ X $ and two stego images $ Z $, $ T $, and get the corresponding image residuals $ Res(X)$, $ Res(Z) $ and $ Res(T) $ separately. And then we calculate the distance between  the residual of cover $ Res(X)$ and the two image residuals $ Res(Z) $ and $ Res(T) $ separately according to a certain $ Dist ( )$ function, denoted as $D_{ZX}$ and $D_{TX}$.  Finally,  we will update the candidate stego $Z$ as the temporary $T$ if $D_{ZX} > D_{TX}$, otherwise we keep $Z$ unchanged.  

We repeat the above operations for all embedding units,  and the whole pseudo-code of the proposed framework  is illustrated in Algothrim \ref{general_pseudo_code}.  The inputs of the algorithm are cover $X$ and the corresponding stego $Y$ using an existing steganography method.  The algorithm first initializes the candidate stego $Z$ as $Y$, and then updates $Z$ using three loops.  In the first loop (i.e. line 6 - 22),  it traverses all the embedding units row by row.  In the second loop (i.e. line 7 - 21), it considers the direction of post-modification to an embedding unit (positive $+$ or negative $-$).  In the third loop (i.e. line 8 - 20), it considers different  amplitudes of post-modification to an embedding units  (e.g. $ 4,  8,  12, \cdots $).  After the three loops, the algorithm  finally outputs a modified stego $Z$, which usually has smaller residual distance compared with the input stego $Y$.  

\begin{algorithm}[!t]
    \caption{Pseudo-code for the  stego post-processing.  $X, Y, Z$  denote the values are of size $n_1\times n_2$.   The for loop in Line 6 traverses all embedding units row by row.}
    \label{general_pseudo_code}
    \begin{algorithmic}[1] 
      \State \textbf{Input: } cover image $ X $;  stego image $ Y $
    	 \State \textbf{Output: } modified stego image $ Z $
	
       \State Initialize $ Z = Y $
       \State $ R_X = Res(X) $
        \State $ R_Z = Res(Z) $
	 	
	 \For { $ i \in \{1, ..., n_1\}, \  j \in \{1, ..., n_2\} $ }
	   \For { $ s \in \{+4, -4\}$ }
	     \While { $ Z_{ij} + s \in \mathcal{V} $ }
	
	     \State $ T = Z $
		\State $ T_{ij} = Z_{ij} + s $		
		\State $ R_T = Res(T) $
		
 		\State $ D_{TX} = Dist(R_T, R_X) $
		\State $ D_{ZX} = Dist(R_Z, R_X) $
		\If{$ D_{TX} < D_{ZX} $ }
\State Update $ Z_{ij} = T_{ij} $	
\State Update $ R_Z = R_T $	
		\Else
		  \State \textbf{break}	
		\EndIf
		\EndWhile
	   \EndFor
	 \EndFor
	 \State \textbf{return} $ Z $
    \end{algorithmic}
\end{algorithm}

\vspace{0.5em}\subsubsection{\textbf{Hyper-Parameters}}
\label{subsub:residual_Distance}

The residual function $ Res( ) $  in  Algorithm \ref{general_pseudo_code} is the key issue  that would significantly affect the security performance of the  proposed framework.  In the following, we will discuss the design of adaptive filter.  

As described in section \ref{subsub:Main_Idea},   most modern steganalytic features are  mainly derived from image residuals.  Thus, the selection of high-pass filters is very important for steganalysis.  Until now,  there are many available filters in existing works,  such as various filters  in  SRM \cite{fridrich2012rich} and GFR \cite{song2015steganalysis}.  
Note that these filters are fixed for all images.  Inspired  from \cite{ker2008revisiting},  we  employ an adaptive way to learn high-pass filters for each image. Specifically,  we first compute the convolution of the image with a prediction filter whose center element is 0, which amounts to predicting target pixels via their surrounding pixels;  and then we determine the elements in the prediction filter by minimizing the mean square error between the predicted pixels and actual ones via lease squares;  finally, we set the center of prediction filter as -1 to obtain the final filter, which calculate the residual between the predicted and actual values.  Different from work \cite{ker2008revisiting} which learns a filter of size $w \times w$ ($w > 2$) with symmetry constraint,  we first learn a base filter of size $1 \times w$ without symmetry constraint for any given image and its transposed version.   Thus,  the resulting basic filter (denoted as $B$) is a  predictor of horizontal direction while its transposed version $B^T$ is  a predictor of vertical direction.  And then we get the outer product of $B$ and $B^T$, denoted as $B \otimes B^T$, which can calculate the residual based on the prediction of both horizontal and vertical directions. 
In our method, we can obtain different residual functions via the combination of  the elements in set $\{B, B^T,  B \otimes B^T\}$ with different size $w$.  Based on our experiments,  we finally select the filters $\{B, B^T, B \otimes B^T\}$ of size 7 for spatial steganography while filters $\{B, B^T\}$ of size 3 for JPEG steganography.  Please note that we take into account several residuals by summing them up. More experimental results on the hyper-parameter selection are shown in  section \ref{subsec:validation}.

In  Algorithm 1 (line 12-13), the distance function $Dist( )$ is used to measure the residual distance between cover and  stego.  Different distances will lead to different post-modification, and thus affect security performance. We have tested several typical distance measures,  including  Manhattan,  Euclidean, Chebychev and  Hamming,  and found that  the Manhattan distance usually  performs well on various steganography methods in both spatial and JPEG domains.  Thus, we employ the  Manhattan distance as follows  in our  experiments.  
\begin{equation}
\label{eq:distance}
d(p, q) = ||p - q||_1 = \sum_{i=1}^{n} |p_i - q_i| 
\end{equation}
where $ p = (p_1, p_2, ... , p_n)$, $ q = (q_1, q_2, ... , q_n)$.

\subsection{Acceleration Strategies}
\label{subsec:Acceleration}

Several issues would significantly affect the processing time of the proposed  Algorithm \ref{general_pseudo_code}.  First of all, there are three nested loops. The first loop will traverse all embedding units of the input stego $Y$. Taking an image of size $512\times 512$ for example, there are totally 262144 ($=512\times 512$)  units to be dealt with.  For each  unit,  two directions (i.e.  positive or negative in  the second loop) and different modification amplitudes (i.e. 4, 8,  $\ldots$,  in the third loop) need to be considered.   If we can reduce the iteration number of these loops,  the algorithm speed will be  improved.  Furthermore,   the  filtering  operation to get image residuals in  the innermost loop  is  time-consuming.   As a result,  a fast method for filtering is  needed.   In the following sections, we will describe  four acceleration strategies separately.  

\vspace{0.5em}\subsubsection{\textbf{Restriction on Position of Post-Modification}}
To reduce the number of embedding units  to be dealt with in the first loop,  we conduct experiment to report the post-modification rate in the set of units modified with steganography. Table \ref{tab:spp_embed_prop_spatial} and  Table \ref{tab:spp_embed_prop_jpeg}  show the average results on 10,000 images from BOSSBase  \cite{bas2011break} for  steganography methods in spatial and JPEG domains.   From Tables  \ref{tab:spp_embed_prop_spatial} and   \ref{tab:spp_embed_prop_jpeg}, we observe that for those embedding units modified by the stego post-processing,  more of them are located at the small set of the units modified with steganography.  Taking  S-UNIWARD for instance, over 65\% of post-modification are located at the set of units modified with steganography, which  occupies only 7.39\% of all  units. Since the steganography modification rate  is relatively lower in the experiment (less than 11\% for  spatial steganography for payload 0.4 bpp,  and less than 4\% for  JPEG steganography for payload 0.4 bpnz), we consider dealing with those embedding units that have been previously modified with the steganography while skipping most unchanged units.  

\begin{table}[!t]
  \renewcommand\arraystretch{1.1}
  \caption{The post-modification rate in the set of units modified with steganography (\%)  evaluated on different spatial steganography methods  (0.4 bpp).}
  \label{tab:spp_embed_prop_spatial}
  \centering
  \begin{tabular}{cccc|c}
     \hline
    \textbf{S-UNI} & \textbf{MIPOD} & \textbf{HILL} & \textbf{CMD-HILL} & \textbf{Average}  \\
    \hline \hline
     65.18 & 66.31 & 65.93 & 65.56 & 65.75 \\
    \hline
  \end{tabular}
\end{table}

\begin{table}[!t]
  \renewcommand\arraystretch{1.1}
  \caption{The post-modification rate in the set of units modified with steganography (\%)  evaluated on different JPEG steganography methods  (0.4 bpnz).}
  \label{tab:spp_embed_prop_jpeg}
  \centering
  \begin{tabular}{c||ccc|c}
     \hline
    \textbf{QF} & \textbf{J-UNI} & \textbf{UERD} & \textbf{BET-HILL} & \textbf{Average} \\ \hline \hline    
    75 & 68.18 & 52.26 & 50.22 & 56.89 \\ \hline   
    95 & 65.30 & 54.67 & 53.70 & 57.89 \\
    \hline
  \end{tabular}
\end{table}

\vspace{0.5em}\subsubsection{\textbf{Restriction on Direction of Post-Modification}}
In the previous section, we  limited the post-modification to be  performed on those embedding units which have been  modified with steganography.  To speed up the second loop, we will analyze the relationship between the directions of  post-modification and steganography modification. We consider the post-modification that locate in the units that has modified by steganography and report the ratio of the post-modifications whose direction is opposite to that of steganography modifications in Table \ref{tab:direction_prop_spatial} and  Table  \ref{tab:direction_prop_jpeg}. The two tables show the average results on 10,000 images from BOSSBase \cite{bas2011break}  in different cases. From  the two tables,  we observe that the  direction of post-modification is usually contrary to that of steganography modification.  On average, such ratio is over 98\% and over 86\% for  spatial and JPEG steganography separately.  In our method,  therefore,  we will limit the direction of post-modification.  This property is reasonable since the detectable artifacts left by steganography usually become  more obvious when the direction of post-modification  is the same as that of steganography modification.

\begin{table}[!t]
  \renewcommand\arraystretch{1.1}
  \caption{The ratio of the post-modifications whose direction is opposite to that of steganography modifications  (\%) evaluated on different spatial steganography methods (payload 0.4 bpp).}
  \label{tab:direction_prop_spatial}
  \centering
  \begin{tabular}{cccc|c}
     \hline
    \textbf{S-UNI} & \textbf{MIPOD} & \textbf{HILL} & \textbf{CMD-HILL}  & \textbf{Average}   \\
    \hline
    \hline   
    99.59 & 98.91 & 99.01 & 96.28 & 98.45 \\
    \hline
  \end{tabular}
\end{table}

\begin{table}[!t]
  \renewcommand\arraystretch{1.1}
  \caption{The ratio of the post-modifications whose direction is opposite to that of steganography modifications (\%) evaluated on different JPEG steganography methods (payload 0.4 bpnz).}
  \label{tab:direction_prop_jpeg}
  \centering
  \begin{tabular}{c||ccc|c}
     \hline
    \textbf{QF} & \textbf{J-UNI} & \textbf{UERD} & \textbf{BET-HILL} & \textbf{Average}   \\ \hline   \hline     
    75 & 95.76 & 92.42 & 89.53 & 92.57 \\ 
    95 & 90.70 & 85.11 & 83.54 & 86.45 \\
     \hline
  \end{tabular}
\end{table}

\vspace{0.5em}\subsubsection{\textbf{Restriction on Amplitude of Post-Modification}} 
In section \ref{subsec:robustness}, we showed that adding a multiple of 4 to any embedding unit of the stego image would not confuse the message extraction. However, most existing literatures have shown that   the  security performance of steganography  usually becomes poorer when the steganography modification becomes relatively larger.   To enhance steganography security,  we expect that  most amplitudes of the post-modification are the smallest ones,  i.e. 4,  for the ternary STCs.  According to our experiments on 10,000 images from BOSSBase \cite{bas2011break}, we observe that the amplitude of 100\% and over 98\% post-modification is equal to 4 for spatial (i.e. S-UNIWARD, MIPOD, HILL, CMD-HILL) and JPEG (i.e. J-UNIWARD, UERD, BET-HILL) steganography methods separately, which fits our expectations very well. Therefore,  we limit the amplitude  for  the post-modification to 4 in our method.

\vspace{0.5em}\subsubsection{\textbf{Efficient Convolution}}
\label{subsubsec:Updating}
In the three previous subsections, we try to reduce the loop count of the three loops in  Algorithm \ref{general_pseudo_code} separately.  In this section, we will speed up the key operation  -  i.e.  the $Res( )$ function to obtain image residual  in the innermost loop (i.e. line 11) in  Algorithm \ref{general_pseudo_code}.   

In section \ref{subsub:residual_Distance},  we determine to employ several adaptive convolution filters with a smaller size (i.e. $w=3$ or $w=7$, which is significantly smaller than the image size $n_1$ and $n_2$, i.e. $n_1=n_2=512$ or $256$ in our experiments) to update image residual of temporary stego $T$.    Please note that the convolution is linear and it just affects a small region of embedding units that around the filter center.  Thus, there is no need to  perform the convolution on the whole temporary  stego $T$ to obtain its residual,  since just an element within  $T$ is different from the candidate stego $Z$ (refer line 9-10 in Algorithm 1).   An equivalent and efficient method is employed in our method.  When the image residual of $Z$ is available (i.e. $R_Z$),  the image residual of $T$ can be calculated based on the following formula: 
\begin{equation}
\begin{aligned}
R_T & =  conv(T, F) = conv(Z\pm \delta_{ij} \times 4, F) \\
& = conv(Z, F) \pm conv(\delta_{ij}, F) \times 4 \\
& =  R_Z \pm conv(\delta_{ij}, F) \times 4
\end{aligned}
\end{equation}
where $ \delta_{ij} $ is a matrix of the same size $n_1\times n_2$, and its elements are all 0 except that the element at position $(i, j)$ is 1.  

Due to the characteristic of matrix $\delta_{ij}$,  it is very fast to get  the $R_T$ via modifying a small region corresponding to the position $(i,j)$ within $R_Z$,  that is,  a  region of size $w\times w$  for spatial steganography or a region of size  $(w+7)\times (w+7)$  \footnote{Note that here: 1). The Inverse Discrete CosineTransform (IDCT) in JPEG decompression is linear. 2)  Modifying a DCT coefficient in JPEG will affect an $8\times 8$ image block in spatial domain.} for JPEG steganography.   By doing so, we can obtain over 500 times acceleration both in spatial and JPEG steganography based on our experiments.   

\subsection{The Proposed Method}

The pseudo-code for the proposed stego post-processing is illustrated in   Algorithm \ref{proposed_pseudo_code}.   
 Note that the source code of  the Algorithm \ref{proposed_pseudo_code} can be available at GitHub.  \footnote{Source codes  are available at:  https://github.com/bolin-chen/universal-spp}
According to the first three acceleration strategies,  we observe that  only one loop (i.e. line 6-20  in Algorithm \ref{proposed_pseudo_code})  is remaining here compared to Algorithm \ref{general_pseudo_code},  and most embedding units in this loops will be skipped (i.e. line 10-12  in   Algorithm \ref{proposed_pseudo_code}).   According to the  analysis  in Section \ref{subsec:Acceleration} - 4),    the execution time is unbearable  without  using the fast method for obtaining temporary image residual .   Thus the fast method is employed in both  Algorithm \ref{general_pseudo_code} and Algorithm \ref{proposed_pseudo_code}.   For a fair comparison,   both algorithms are implemented with Matlab and  on the same server with CPU Intel Xeon Gold 6130.  
The processing time and the security performance of two algorithms would be evaluated in the following. 

\begin{algorithm}[!t]
    \caption{Pseudo-code for the  stego post-processing.  Images $X, Y, Z$ are of size $n_1\times n_2$.  $ F $ is the adaptive filter generated based on predictor learned from $ X $. The for loop in Line 6 traverses all embedding units row by row.}
    \label{proposed_pseudo_code}
    \begin{algorithmic}[1] 
      \State \textbf{Input: } cover image $ X $; stego image $ Y $
    	 \State \textbf{Output: } enhanced stego images $ Z $
       \State Initialize $ Z = Y $
       \State $ R_X = conv(X, F) $
	  \State $ R_Z = conv(Z, F) $	
	 \For { $ i \in \{1, ..., n_1\}, \  j \in \{1, ..., n_2\} $ }
		\State $ s  = (X_{ij} - Y_{ij}) \times 4 $
	        \State $ T = Z $
	        \State $ T_{ij} = Z_{ij} +  s$
	     \If { $ s == 0$ \textbf{or}  $ T_{ij}  \notin \mathcal{V} $ }
	     \State continue
	     \EndIf
		\State $ R_T = $  Update a small region of $ R_Z $			
	 	\State $ D_{ZX} = \sum |R_Z - R_X| $
		\State $ D_{TX} = \sum |R_T - R_X| $
		\If{$ D_{TX} < D_{ZX} $ }
			\State Update $ Z_{ij} = T_{ij}$
			\State Update $ R_Z = R_T $	
		\EndIf
	 \EndFor
	 \State \textbf{return} $ Z $
    \end{algorithmic}
\end{algorithm}

\vspace{0.5em}\subsubsection{\textbf{Comparison on Processing Time}}
In this experiment, we will compare the processing time of  the algorithm before and after using the first three  acceleration strategies.  Four spatial  steganography methods  and three  JPEG steganograpy methods are considered  \footnote{Please refer to Section  \ref{sec:experiments} for more details about the experimental settings.}.  The comparative results  are shown  in Table \ref{tab:filter_all_spatial_time} and Table \ref{tab:filter_all_jpeg_time}.   From the two tables, 
 we  observe that  the processing time of the proposed method (i.e. Algorithm 2)  is significantly shorter  than the original one (i.e. Algorithm 1).  On average, we gain  over 12  and 9 times speed improvement for the spatial and JPEG steganography separately.

\begin{table}[!t]
  \renewcommand\arraystretch{1.1}
  \caption{Processing time (s)  for  Algorithm 1 and Algorithm 2 in spatial domain  (0.4 bpp).  }
  \label{tab:filter_all_spatial_time}
  \centering
  \begin{tabular}{c||cccc|c}
     \hline
     \textbf{} & \textbf{S-UNI} & \textbf{MIPOD} & \textbf{HILL} & \textbf{CMD-HILL}  & \textbf{Average} \\
    \hline
    \hline 
    Algorithm 1       & 3.75 & 3.72 & 3.75 & 3.72 & 3.74 \\
    Algorithm 2          & 0.27 & 0.29 & 0.29 & 0.33 & 0.30 \\
    \hline
  \end{tabular}
\end{table}

\begin{table}[!t]
  \renewcommand\arraystretch{1.1}
  \caption{Processing time (s)  for  Algorithm 1 and Algorithm 2 in JPEG domain  (0.4 bpnz). }
  \label{tab:filter_all_jpeg_time}
  \centering
  \begin{tabular}{c||c||ccc|c}
     \hline
    \textbf{QF} & \textbf{Strategy} & \textbf{J-UNI} & \textbf{UERD} & \textbf{BET-HILL} & \textbf{Average} \\
    \hline
    \hline
    \multirow{2}{*}{75} & Algorithm 1     & 4.65 & 4.66 & 4.64 & 4.65 \\
    & Algorithm 2  & 0.50 & 0.49 & 0.49 & 0.49 \\
    \hline
    \multirow{2}{*}{95} & Algorithm 1       & 4.83 & 4.85 & 4.81 & 4.83 \\
    & Algorithm 2  & 0.54 & 0.53 & 0.53 & 0.53 \\
    \hline
  \end{tabular}
\end{table}

\vspace{0.5em} \subsubsection{\textbf{Comparison on Security Performance}}

In this experiment, we will compare the  security performances of  Algorithm 1 and Algorithm 2.   The experimental results are shown in Table \ref{tab:filter_all_spatial} and Table \ref{tab:filter_all_jpeg}.   From the two tables,  we observe that both algorithms can enhance the steganography security in all cases.  Although we have significantly simplified the Algorithm 1 for acceleration,   the performance of Algorithm 2 would not drop.  On the contrary,  it is able to outperform Algorithm 1 slightly on average.

\begin{table}[!t]
  \renewcommand\arraystretch{1.1}
  \caption{Detection accuracies (\%)  for Algorithm 1 and Algorithm 2  in spatial domain (0.4 bpp).  The steganalytic feature set is SRM.  In all following tables, the value with an asterisk (*) denotes the best result.}
  \label{tab:filter_all_spatial}
  \centering
  \begin{tabular}{c||cccc|c}
     \hline
     \textbf{} & \textbf{S-UNI} & \textbf{MIPOD} & \textbf{HILL} & \textbf{CMD-HILL}  & \textbf{Average} \\
    \hline
    \hline   
    Baseline   & 79.68 & 75.66 & 75.72 & 70.06 & 75.28 \\
    \hline     
  Algorithm 1     & \textblue{78.34} & \textblue{72.83*} & \textblue{72.96} & \textblue{69.35} & \textblue{73.37} \\    
  Algorithm 2          & \textblue{78.31*} & \textblue{73.19} & \textblue{72.42*} & \textblue{68.93*} & \textblue{73.21*} \\
    \hline
  \end{tabular}
\end{table}

\begin{table}[!t]
  \renewcommand\arraystretch{1.1}
  \caption{Detection accuracies (\%)  for Algorithm 1 and Algorithm 2  in JPEG domain (0.4 bpnz).  The steganalytic feature set is GFR.}
  \label{tab:filter_all_jpeg}
  \centering
  \begin{tabular}{c||c||ccc|c}
     \hline
    \textbf{QF} & \textbf{Strategy} & \textbf{J-UNI} & \textbf{UERD} & \textbf{BET-HILL} & \textbf{Average} \\
    \hline
    \hline
    \multirow{3}{*}{75} & Baseline   & 89.66 & 89.59 & 87.13  & 88.79 \\
    \cline{2-6}
    & Algorithm 1    & \textblue{88.81} & \textblue{88.15} & \textblue{84.76} & \textblue{87.24} \\
    & Algorithm 2  & \textblue{88.54*} & \textblue{87.48*} & \textblue{84.59*} & \textblue{86.87*} \\
    \hline
    \multirow{3}{*}{95} & Baseline   & 72.79 & 76.00 & 69.22 & 72.67 \\
    \cline{2-6}
    & Algorithm 1      & \textblue{72.26} & \textblue{73.72} & \textblue{66.31*} & \textblue{70.76} \\
    & Algorithm 2  & \textblue{71.75*} & \textblue{73.46*} & \textblue{66.88} & \textblue{70.70*} \\  
    \hline
  \end{tabular}
\end{table}

\section{Experimental Results and Discussions}
\label{sec:experiments}

In our experiments,  we collect 10,000 gray-scale images of size $512\times 512$ from BOSSBase \cite{bas2011break}, and  randomly divide them into two non-overlapping and equal parts, one for training and the other for testing.  Like most existing literatures, we use the optimal simulator for data embedding.  Four typical spatial steganography methods (i.e.  S-UNIWARD \cite{holub2014universal}, MIPOD \cite{sedighi2016content}, HILL \cite{li2014new} and CMD-HILL \cite{li2015strategy})  and three typical JPEG steganography methods (i.e. J-UNIWARD \cite{holub2014universal}, UERD \cite{guo2015using} and BET-HILL \cite{hu2018efficient}) are considered.  The spatial steganalytic detectors include two conventional feature sets (i.e. SRM \cite{fridrich2012rich}, maxSRM \cite{denemark2014selection}) and  a CNN-based one (i.e. Xu-Net \cite{xu2016structural}). Similarly, the JPEG steganalytic detectors also include two  conventional ones (i.e. GFR \cite{song2015steganalysis}, SCA-GFR \cite{denemark2016steganalysis} ) and a CNN-based one (i.e. J-Xu-Net \cite{xu2017deep}).  The ensemble classifier \cite{kodovsky2012ensemble} is used for conventional steganalytic features.

\subsection{Hyper-Parameter Selection} 
\label{subsec:validation}

The residual function $Res( )$ is the key issue  in the proposed algorithm that will significantly affect the security performance.  We employ several adaptive filters to get image residuals.  In this section, we try to select proper hyper-parameter about the adaptive filters,  including  the adaptive filter set and the size of basic filter $B$. 

\begin{table}[!t]
  \renewcommand\arraystretch{1.1}
  \caption{Detection accuracies (\%) for different filter sets in  spatial steganography  (0.4 bpp).  In all following tables,  the underlined value denotes the pooper result compared to the baseline steganography.}
  \label{tab:filter_pattern_spatial}
  \centering
  \begin{tabular}{c||cccc|c}
     \hline 
      \textbf{Filter Set} & \textbf{S-UNI} & \textbf{MIPOD} & \textbf{HILL} & \textbf{CMD-HILL}  & \textbf{Average} \\
    \hline
    \hline   
    Baseline   & 79.68 & 75.66 & 75.72 & 70.06 & 75.28 \\
    \hline   
    $\{ B \otimes B^T \}$      & \textblue{78.82*} & \textblue{74.58} & \textblue{73.87} & \textred{\underline{70.37}} & \textblue{74.41} \\    
    $\{B, B^T\} $          & \textblue{79.42} & \textblue{75.69} & \textblue{75.13} & \textblue{70.02} & \textblue{75.07} \\
    $\{B, B^T, B \otimes B^T \}$     & \textblue{79.16} & \textblue{74.37*} & \textblue{73.75*} & \textblue{69.95*} & \textblue{74.31*} \\
    \hline
  \end{tabular}
\end{table}

\begin{table}[!t]
  \renewcommand\arraystretch{1.1}
  \caption{Detection accuracies (\%) for different filter sets in  JPEG  steganography  (0.4 bpnz). }
   \label{tab:filter_pattern_jpeg}
  \centering
  \begin{tabular}{c||c||ccc|c}
     \hline
    \textbf{QF} & \textbf{Filter Set} & \textbf{J-UNI} & \textbf{UERD} & \textbf{BET-HILL}  & \textbf{Average} \\
    \hline
    \hline
    \multirow{4}{*}{75} & Baseline   & 89.66 & 89.59 & 87.13  & 88.79 \\
    \cline{2-6}
    & $\{B \otimes B^T\}$       & \textred{\underline{94.97}} & \textred{\underline{95.43}} & \textred{\underline{92.99}} & \textred{\underline{94.46}} \\
    & $\{B, B^T\}$           & \textblue{88.54*} & \textblue{87.48*} & \textblue{84.59} & \textblue{86.87*} \\
    & $\{B, B^T, B \otimes B^T \}$  & \textblue{88.56} & \textblue{87.67} & \textblue{84.54*} & \textblue{86.92} \\
    \hline
    \multirow{4}{*}{95} & Baseline   & 72.79 & 76.00 & 69.22 & 72.67 \\
    \cline{2-6}    
    & $\{B \otimes B^T\}$       & \textred{\underline{80.56}} & \textred{\underline{85.42}} & \textred{\underline{75.81}} & \textred{\underline{80.60}} \\    
    &$\{B, B^T\}$            & \textblue{71.75*} & \textblue{73.46*} & \textblue{66.88} & \textblue{70.70*} \\
    & $\{B, B^T, B \otimes B^T \}$ & \textblue{72.09} & \textblue{73.80} & \textblue{66.67*} & \textblue{70.85} \\
    \hline
  \end{tabular}
\end{table}

\begin{table}[!t]
  \renewcommand\arraystretch{1.1}
  \caption{Detection accuracies (\%)  for different sizes of basic filter $B$ for 
 spatial steganography (0.4 bpp).  The  filter set  is  $\{B, B^T, B \otimes B^T \}$.  The steganalytic feature set is SRM. }
  \label{tab:filter_length_spatial}
  \centering
  \begin{tabular}{c||cccc|c}
     \hline

     \textbf{Size} & \textbf{S-UNI} & \textbf{MIPOD} & \textbf{HILL} & \textbf{CMD-HILL}  & \textbf{Average} \\

    \hline
    \hline

    Baseline   & 79.68 & 75.66 & 75.72 & 70.06 & 75.28 \\
    \hline    
    
    3           & \textblue{79.16} & \textblue{74.37} & \textblue{73.75} & \textblue{69.95} & \textblue{74.31} \\
    5       & \textblue{78.28*} & \textblue{73.32} & \textblue{72.46} & \textblue{69.33} & \textblue{73.35} \\
    7  & \textblue{78.31} & \textblue{73.19*} & \textblue{72.42*} & \textblue{68.93*} & \textblue{73.21*} \\
    9  & \textblue{78.37} & \textblue{73.29} & \textblue{73.15} & \textblue{69.22} & \textblue{73.51} \\
    \hline

  \end{tabular}
\end{table}

\begin{table}[!t]
  \renewcommand\arraystretch{1.1}
  \caption{{Detection accuracies (\%)  for different sizes of basic filter $B$ for 
 JPEG steganography (0.4 bpnz).  The  filter set  is   $\{B, B^T\}$.  The steganalytic feature set is GFR.  }}
  \label{tab:filter_length_jpeg}
  \centering
  \begin{tabular}{c||c||ccc|c}
     \hline

     \textbf{QF} & \textbf{Size} & \textbf{J-UNI} & \textbf{UERD} & \textbf{BET-HILL}  & \textbf{Average} \\

    \hline
    \hline

    \multirow{5}{*}{75} & Baseline   & 89.66 & 89.59 & 87.13  & 88.79 \\
    \cline{2-6}

    & 3  & \textblue{88.54*} & \textblue{87.48*} & \textblue{84.59*} & \textblue{86.87*} \\
    & 5  & \textblue{88.87} & \textblue{87.82} & \textblue{84.95} & \textblue{87.21} \\
    & 7  & \textblue{88.81} & \textblue{87.99} & \textblue{85.35} & \textblue{87.38} \\
    & 9  & \textblue{88.89} & \textblue{88.06} & \textblue{85.33} & \textblue{87.43} \\

    \hline

    \multirow{5}{*}{95} & Baseline   & 72.79 & 76.00 & 69.22 & 72.67 \\
    \cline{2-6}

    & 3  & \textblue{71.75*} & \textblue{73.46*} & \textblue{66.88} & \textblue{70.70*} \\
    & 5  & \textblue{72.09} & \textblue{73.92} & \textblue{66.81*} & \textblue{70.94} \\
    & 7  & \textblue{72.04} & \textblue{73.99} & \textblue{66.90} & \textblue{70.98} \\
    & 9  & \textblue{72.34} & \textblue{74.17} & \textblue{66.90} & \textblue{71.14} \\
    \hline

  \end{tabular}
\end{table}

\begin{table*}[!]
  \renewcommand\arraystretch{1.1}
  \setlength{\tabcolsep}{4pt}
  \caption{Detection accuracies (\%) for different steganography methods in spatial domain.  In all following tables, we name the enhanced version of some steganography such as ``A''  with the proposed Stego Post-Processing as ``A-SPP'' for short. }
  \label{tab:security_spatial}
  \centering
  \begin{tabular}{c||ccccc||ccccc||ccccc}
     \hline

    \multirow{2}{*}{\textbf{Steganography}} & \multicolumn{5}{c||}{\textbf{SRM}} & \multicolumn{5}{c||}{\textbf{maxSRMd2}}  & \multicolumn{5}{c}{\textbf{Xu-Net}} \\

    \cline{2-16}

    & \textbf{0.1} & \textbf{0.2} & \textbf{0.3} & \textbf{0.4} & \textbf{0.5} & \textbf{0.1} & \textbf{0.2} & \textbf{0.3} & \textbf{0.4} & \textbf{0.5} & \textbf{0.1} & \textbf{0.2} & \textbf{0.3} & \textbf{0.4} & \textbf{0.5} \\

    \hline
    \hline
    S-UNI             & 60.09 & 68.29 & 74.74 & 79.68 & 83.61 & 63.72 & 70.71 & 76.58 & 80.69 & 84.36 & 55.72 & 64.86 & 73.62 & 78.60 & 82.72 \\
    S-UNI-SPP         & \textblue{59.76*} & \textblue{67.55*} & \textblue{73.27*} & \textblue{78.31*} & \textblue{82.22*} & \textblue{63.31*} & \textblue{70.09*} & \textblue{75.16*} & \textblue{79.17*} & \textblue{82.75*} & \textblue{55.24*} & \textblue{63.71*} & \textblue{70.66*} & \textblue{75.37*} & \textblue{79.73*} \\
    \hline
    MIPOD           & 58.25 & 65.68 & 71.44 & 75.66 & 80.20 & 60.77 & 67.37 & 72.92 & 77.41 & 81.34 & 58.06 & 65.52 & 71.11 & 75.66 & 80.43 \\
    MIPOD-SPP       & \textred{\underline{58.37}} & \textblue{63.83*} & \textblue{69.11*} & \textblue{73.19*} & \textblue{77.52*} & \textblue{59.36*} & \textblue{65.21*} & \textblue{70.15*} & \textblue{73.79*} & \textblue{78.33*} & \textblue{56.98*} & \textblue{62.38*} & \textblue{66.80*} & \textblue{71.36*} & \textblue{75.23*} \\
    \hline
    HILL            & 56.65 & 64.14 & 70.44 & 75.72 & 79.67 & 62.43 & 69.32 & 73.72 & 78.30 & 81.92 & 58.04 & 65.50 & 71.40 & 77.26 & 80.23 \\
    HILL-SPP        & \textblue{56.09*} & \textblue{62.60*} & \textblue{67.68*} & \textblue{72.42*} & \textblue{76.47*} & \textblue{60.82*} & \textblue{66.82*} & \textblue{71.48*} & \textblue{75.72*} & \textblue{79.23*} & \textblue{56.29*} & \textblue{62.08*} & \textblue{66.92*} & \textblue{71.36*} & \textblue{75.50*} \\
    \hline
    CMD-HILL        & 55.09 & 60.53 & 65.86 & 70.06 & 74.41 & 59.79 & 65.54 & 69.74 & 73.35 & 76.46 & 54.81 & 60.19 & 64.66 & 69.64 & 73.39 \\
    CMD-HILL-SPP    & \textblue{54.55*} & \textblue{60.13*} & \textblue{64.95*} & \textblue{68.93*} & \textblue{72.73*} & \textblue{59.40*} & \textblue{64.42*} & \textblue{68.81*} & \textblue{71.83*} & \textblue{75.50*} & \textblue{54.39*} & \textblue{59.07*} & \textblue{62.65*} & \textblue{67.44*} & \textblue{70.25*} \\
    \hline

  \end{tabular}
\end{table*}

\begin{table*}[!]
  \renewcommand\arraystretch{1.1}
  \setlength{\tabcolsep}{4pt}
  \caption{Detection accuracies (\%) for different steganography methods in JPEG domain. }
  \label{tab:security_jpeg}
  \centering
  \begin{tabular}{c||c||ccccc||ccccc||ccccc}
     \hline

    \multirow{2}{*}{\textbf{QF}} & \multirow{2}{*}{\textbf{Steganography}} & \multicolumn{5}{c||}{\textbf{GFR}} & \multicolumn{5}{c||}{\textbf{SCA-GFR}} & \multicolumn{5}{c}{\textbf{J-Xu-Net}} \\

    \cline{3-17}

    & & \textbf{0.1} & \textbf{0.2} & \textbf{0.3} & \textbf{0.4} & \textbf{0.5} & \textbf{0.1} & \textbf{0.2} & \textbf{0.3} & \textbf{0.4} & \textbf{0.5} & \textbf{0.1} & \textbf{0.2} & \textbf{0.3} & \textbf{0.4} & \textbf{0.5} \\

    \hline
    \hline
    \multirow{6}{*}{75} & J-UNI             & 59.03 & 71.00 & 81.82 & 89.66 & 94.50 & 64.33 & 76.91 & 85.94 & 91.75 & 95.47 & 65.28 & 77.66 & 86.13 & 91.72 & 95.01 \\
    & J-UNI-SPP         & \textred{\underline{59.06}} & \textblue{70.92*} & \textblue{81.23*} & \textblue{88.54*} & \textblue{93.46*} & \textblue{63.72*} & \textblue{76.36*} & \textblue{85.07*} & \textblue{90.87*} & \textblue{94.65*} & \textblue{\underline{65.23}} & \textblue{77.53*} & \textblue{85.97*} & \textblue{91.46*} & \textblue{94.57*} \\
    \cline{2-17}
    & UERD           & 60.42 & 72.46 & 82.27 & 89.59 & 94.14 & 70.36 & 82.17 & 88.91 & 93.17 & 95.88 & 77.44 & 88.04 & 93.01 & 96.13 & 97.46 \\
    & UERD-SPP       & \textblue{59.60*} & \textblue{71.52*} & \textblue{81.11*} & \textblue{87.48*} & \textblue{92.54*} & \textblue{70.07*} & \textblue{81.08*} & \textblue{88.04*} & \textblue{92.10*} & \textblue{94.83*} & \textblue{77.06*} & \textred{\underline{88.18}} & \textblue{92.58*} & \textblue{95.95*} & \textred{\underline{97.58}} \\
    \cline{2-17}
    & BET-HILL            & 58.26 & 69.12 & 78.96 & 87.13 & 92.10 & 65.19 & 76.98 & 86.11 & 92.01 & 95.58 & 65.63 & 77.70 & 84.88 & 90.43 & 95.20 \\
    & BET-HILL-SPP        & \textblue{57.82*} & \textblue{67.58*} & \textblue{77.04*} & \textblue{84.59*} & \textblue{90.42*} & \textblue{64.08*} & \textblue{75.22*} & \textblue{83.85*} & \textblue{89.73*} & \textblue{93.59*} & \textblue{64.28*} & \textblue{76.62*} & \textblue{83.27*} & \textblue{89.57*} & \textblue{93.73*} \\

    \hline

    \multirow{6}{*}{95} & J-UNI             & 52.41 & 57.92 & 65.15 & 72.79 & 80.63 & 53.59 & 59.94 & 67.21 & 73.90 & 80.00 & 50.26 & 57.88 & 66.43 & 73.38 & 79.03 \\
    & J-UNI-SPP         & \textblue{52.31*} & \textblue{57.66*} & \textblue{64.55*} & \textblue{71.75*} & \textblue{78.98*} & \textblue{53.52*} & \textblue{59.47*} & \textblue{65.98*} & \textblue{72.65*} & \textblue{78.27*} & \textblue{50.08*} & \textblue{57.72*} & \textblue{65.34*} & \textblue{72.42*} & \textred{\underline{79.18}} \\
    \cline{2-17}
    & UERD           & 54.18 & 60.62 & 68.49 & 76.00 & 82.77 & 59.33 & 67.89 & 74.57 & 80.53 & 85.44 & 50.12 & 73.37& 82.39 & 88.97 & 92.79 \\
    & UERD-SPP       & \textblue{54.11*} & \textblue{60.01*} & \textblue{66.66*} & \textblue{73.46*} & \textblue{79.68*} & \textblue{59.06*} & \textblue{66.81*} & \textblue{72.85*} & \textblue{77.93*} & \textblue{82.65*} & \textblue{50.04*} & \textblue{72.74*} & \textblue{82.30*} & \textblue{88.17*} & \textblue{92.12*} \\
    \cline{2-17}
    & BET-HILL            & 52.24 & 56.75 & 62.30 & 69.22 & 75.59 & 54.14 & 59.47 & 65.36 & 71.73 & 77.81 & 50.47 & 58.49 & 65.36 & 73.01 & 80.00 \\
    & BET-HILL-SPP        & \textblue{52.06*} & \textblue{56.21*} & \textblue{61.22*} & \textblue{66.88*} & \textblue{72.86*} & \textblue{53.42*} & \textblue{58.19*} & \textblue{63.39*} & \textblue{69.18*} & \textblue{75.04*} & \textblue{49.90*} & \textblue{56.58*} & \textblue{64.09*} & \textblue{72.60*} & \textblue{78.07*} \\
    \hline
  \end{tabular}
\end{table*}

\vspace{0.5em}\subsubsection{\textbf{Adaptive Filter Set}}
As described in section \ref{subsub:residual_Distance},  we first learn a basic filter $B$ for each image, and then produce  two filters via transpose and outer product, and then we obtain three adaptive filters,  that is,  $B$,  $B^T$  and $B \otimes B^T$.   
For simplification,  three  combinations of above filters are evaluated, that is, $\{B \otimes B^T\}$,  $\{B, B^T\}$ and  $\{B, B^T, B \otimes B^T\}$.  In addition,  the filter size of $B$ is fixed as 3 in this experiment, and the steganalytic features SRM and GFR are used for security evaluation for the spatial (0.4 bpp) and JPEG steganography (0.4 bpnz) separately. 
The  detection accuracies evaluated on test set are shown in Table  \ref{tab:filter_pattern_spatial} and Table \ref{tab:filter_pattern_jpeg}.  From Table \ref{tab:filter_pattern_spatial}, we observe that the three filter sets can improve the security performance of the four spatial steganography methods except using the filter  $\{B \otimes B^T\}$ on CMD-HILL.  On average, the set  $\{B, B^T, B \otimes B^T \}$  achieves the best performance,  and it gains an  average improvement of 0.97\% compared to the baseline steganpgraphy.  From Table \ref{tab:filter_pattern_jpeg}, we  observe  $\{B, B^T \}$ and  $\{B, B^T, B \otimes B^T \}$  can improve the security performance while  $\{B \otimes B^T \}$  will  significantly drop  the performance.   On average,  the filter set $\{B, B^T \}$ performs the best and it achieves an  improvement of  around 1.90\% for both quality factors.  
The above  results show that different adaptive filter sets  have a great influence on security performance.  The filter sets $\{B, B^T, B \otimes B^T \}$  and  $\{B, B^T\}$ usually perform the best in spatial and JPEG domain separately. 

\vspace{0.5em}\subsubsection{\textbf{Size of  Basic Filter $B$}}
In previous section,  we fixed the size of basic filter $B$ as 3, and selected the proper filter set for spatial and JPEG steganography separately.  In this section,  we first fixed the selected filter set, and  evaluate their performances with different  sizes of the basic filter $B$,  including $w= 3,  5,  7,  9$.  The  detection accuracies are shown in Table \ref{tab:filter_length_spatial} and Table \ref{tab:filter_length_jpeg}.  From the two tables, we observe that  the four filter  sizes can improve the performance of  various steganography methods in both spatial and JPEG domains.  In spatial domain,  the average performance becomes the best  when the size of $B$ is 7 instead of 3,  which will further gain an improvement of 1.10\%.   In JPEG domain,  the proper size of $B$ is still 3 based on our experiments.  

We should note that the hyper-parameter determined  previously is just a suboptimal solution.  Due to time constraint,  
we probably find a  better solution  via  brute force method according to several important issues, such as the combinations of adaptive filters with different sizes,  the specific steganography with a given payload,  and the  steganalytic models under investigation and so on.  For simplicity,  we just apply  the filter set $\{B, B^T, B \otimes B^T \}$ with filter size $w=7$ for  spatial steganography, and the filter set $\{B, B^T \}$  with filter size $w=3$ for JPEG steganography for all embedding payloads and steganalytic models in the following section.

\subsection{Steganography Security Evaluation}
\label{subsec:bossbase_security}

In this section,  we will evaluate the security performance on  different steganography methods for different payloads ranging from 0.1  bpp/bpnz to 0.5 bpp/bpnz.   Three different steganalyzers in spatial domain,  including SRM  \cite{fridrich2012rich}, maxSRM \cite{denemark2014selection}, and Xu-Net \cite{xu2016structural},  and three steganalyzers in JPEG domain, including GFR  \cite{song2015steganalysis}, SCA-GFR \cite{denemark2016steganalysis},  and J-Xu-Net \cite{xu2017deep}, are used for security evaluation.  The  average detection accuracies on test set are shown in Table \ref{tab:security_spatial} and Table \ref{tab:security_jpeg}.  From the two tables, we  obtain the three following observations: 
\begin{itemize}
\item  Almost  in all cases,  the proposed method can effectively improve the steganography security both in spatial and JPEG  domains. The improvement usually increases with increasing embedding payload.  

\item In spatial domain,   we can achieve greater  improvements on MIPOD and HILL  compared to S-UNIWARD and CMD-HILL. Taking the  payload of 0.5 bpp for instance, we obtain about 3\% improvement on both MIPOD and HILL,   while less than 2\% for two other steganography methods under two hand-crafted  steganalytic feature sets,  i.e.  SRM and maxSRM.   Furthermore, the proposed method seems more effective to  the CNN-based steganalyzer (i.e.  Xu-Net).  For instance, we can obtain about 5\% improvement for MIPOD and HILL for the payload of 0.5 bpp, which is a significant improvement on current steganography methods.     

\begin{figure*}[!]
  \centering
  \begin{subfigure}{0.5\textwidth}
    \centering
    \includegraphics[width=0.975\linewidth]{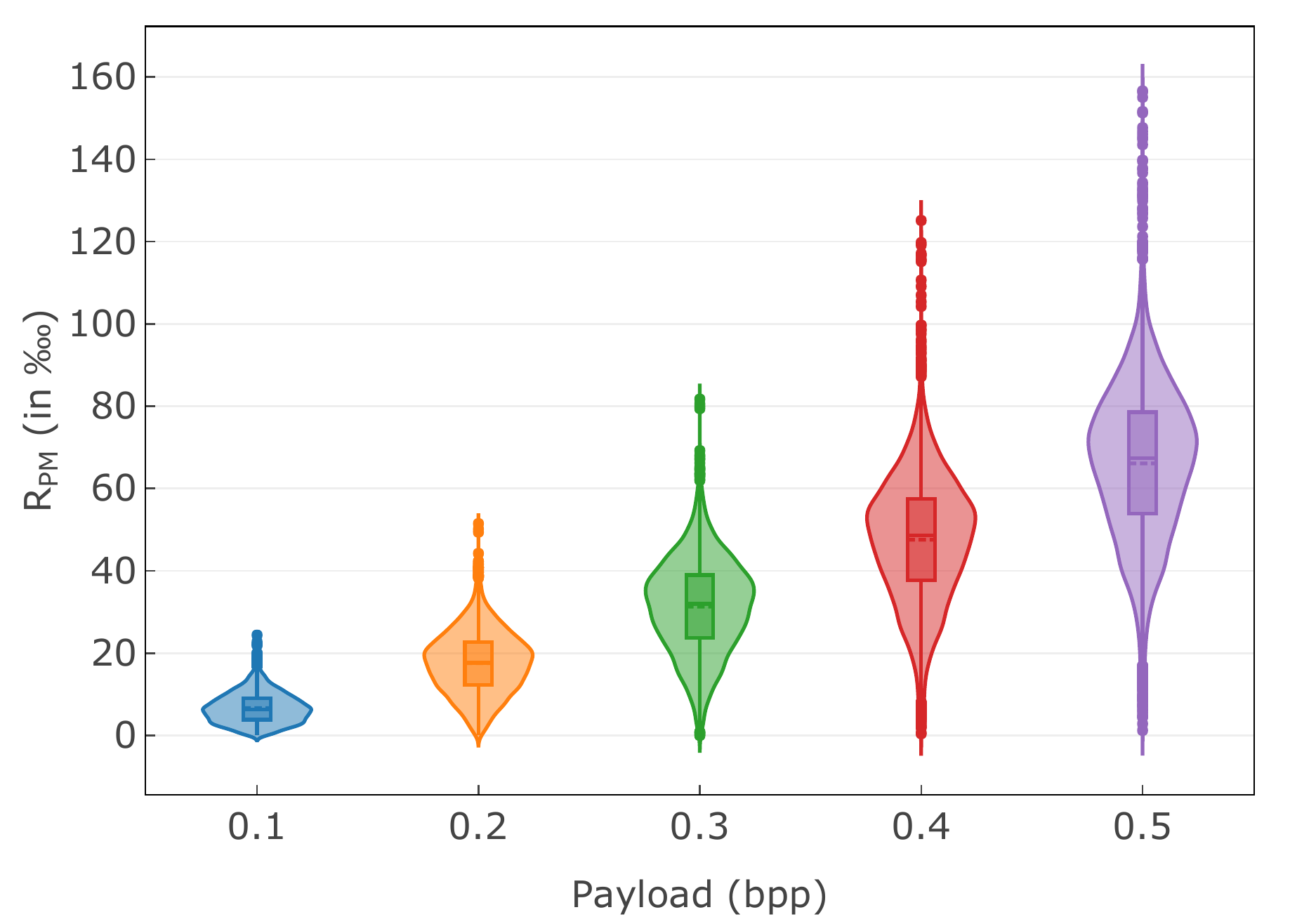}
    \caption{HILL}
   \end{subfigure}%
  \begin{subfigure}{0.5\textwidth}
    \centering
    \includegraphics[width=1\linewidth]{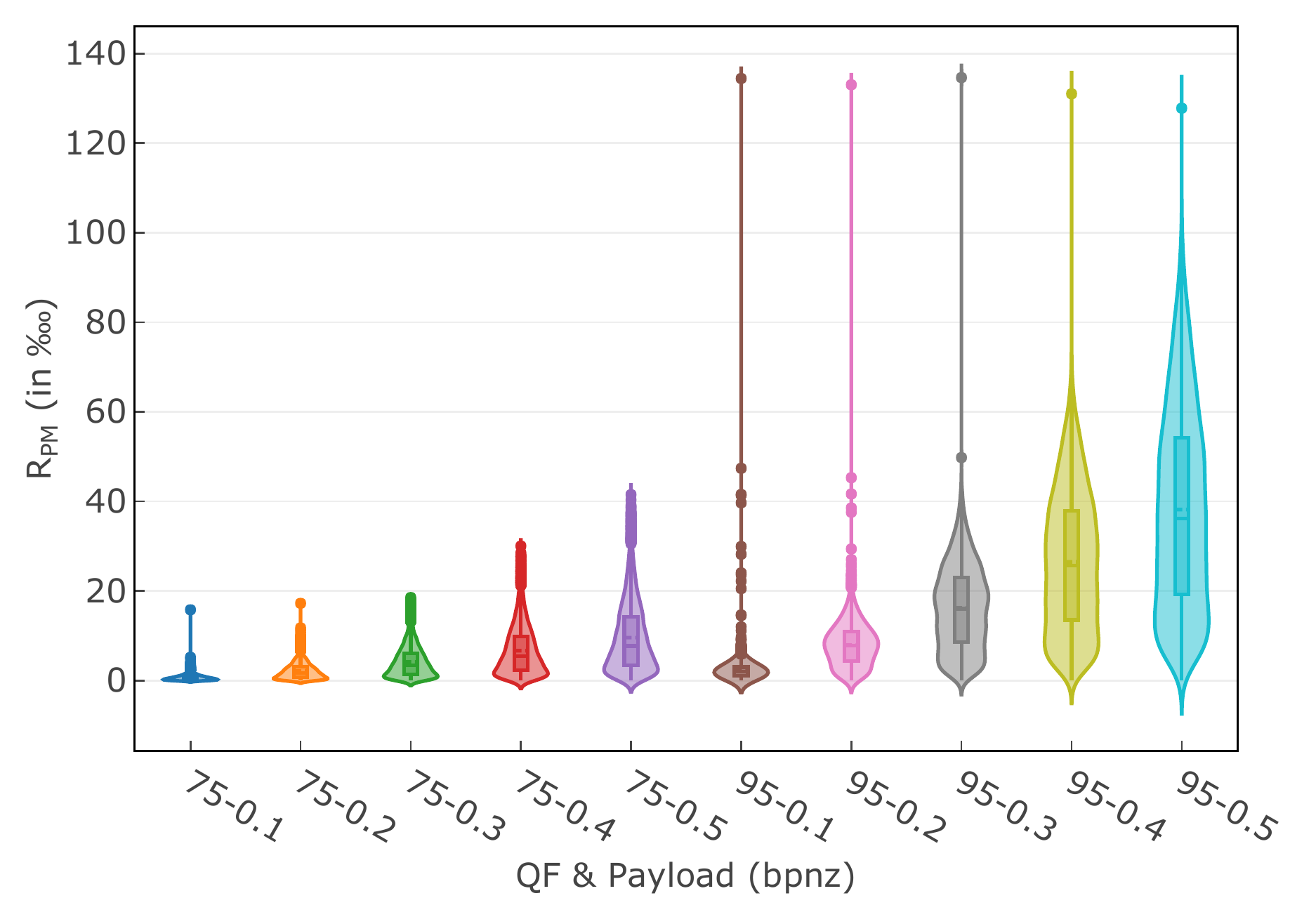}
    \caption{BET-HILL}
  \end{subfigure}%
  \caption{The violin plots of the  Post-Modification Rates for HILL and BET-HILL}
  \label{fig:violin_plot}
\end{figure*}

\item In JPEG domain,  the proposed method can gain more improvement on  UERD and BET-HILL compared to J-UNIWARD.  Taking the payload 0.5 bpnz  and $QF = 95$ for instance,  it obtain an improvement of about 3\% for both UERD and BET-HILL under the steganalytic feature GFR, while only 1.65\% for J-UNIWARD.  In addition, the proposed method seems less effective to CNN-based steganalyzer (i.e. J-Xu-Net) compared to  the hand-crafted  feature sets.  In some cases, the security  will drop slightly (less than 0.15\%) after using the proposed method.  

\end{itemize}

\subsection{Analysis on Post-Modification}
\label{subsec:Post_Modifications}

In this section, we will analyze some statistical characteristics on the post-modification with our method,  including the  modification rate and its relation to the density of steganography modification.

\begin{table}[!]
  \renewcommand\arraystretch{1.1}
  \caption{Post-Modification Rate (\textpertenthousand) for  steganography methods in spatial domain. }
  \label{tab:modification_rate_spatial}
  \centering
  \begin{tabular}{c||ccccc}
     \hline
    \textbf{Steganography} & \textbf{0.1} & \textbf{0.2} & \textbf{0.3} & \textbf{0.4} & \textbf{0.5}\\
    \hline
    \hline
    S-UNI            & 3.81 & 11.21 & 21.30 & 33.92 & 48.89 \\ 
    \hline
    MIPOD            & 3.26 & 12.87 & 27.14 & 45.17 & 66.34 \\
    \hline
    HILL             & 6.64 & 17.53 & 31.30 & 47.54 & 66.09 \\
    \hline
    CMD-HILL         & 2.67 & 7.44 & 13.81 & 21.91 & 31.78 \\
    \hline
  \end{tabular}
\end{table}

\begin{table}[!]
  \renewcommand\arraystretch{1.1}
  \caption{Post-Modification Rate (\textpertenthousand) for  steganography methods in JPEG domain. }
  \label{tab:modification_rate_jpeg}
  \centering
  \begin{tabular}{c||c||ccccc}
     \hline
    \textbf{QF} & \textbf{Steganography} & \textbf{0.1} & \textbf{0.2} & \textbf{0.3} & \textbf{0.4} & \textbf{0.5} \\  \hline \hline
    \multirow{3}{*}{75} & J-UNI & 0.37 & 1.42 & 3.07 & 5.21 & 7.87 \\
    \cline{2-7}
    & UERD & 0.39 & 1.52 & 3.28 & 5.59 & 8.41 \\
    \cline{2-7}
    & BET-HILL & 0.66 & 2.11 & 4.19 & 6.64 & 9.59 \\
    \hline
    \multirow{3}{*}{95}   & J-UNI & 1.33 & 5.73 & 13.13 & 23.00 & 34.74 \\
    \cline{2-7}
    & UERD & 1.84 & 7.16 & 15.48 & 26.09 & 38.48 \\
    \cline{2-7}
    & BET-HILL & 2.21 & 7.87 & 16.24 & 26.53 & 38.23 \\
    \hline
  \end{tabular}
\end{table}

\vspace{0.5em}\subsubsection{\textbf{Post-Modification Rate}}

We define the post-modification rate as follows: 
\begin{equation}
\label{eq:pmr}
R_{PM} = \frac{ |Z \not= Y|}{|Y|} =  \frac{ |Z \not= Y|}{|X|} 
\end{equation}
where $X, Y, Z$ denote the set of  embedding units in cover, stego, and the modified version with the proposed method separately. Note that $|X| = |Y| = |Z|$  since the number of  embedding units is the same for the three images. Table \ref{tab:modification_rate_spatial} and  Table \ref{tab:modification_rate_jpeg} show the 
average results evaluated on 10,000 images from BossBase. From the two tables, we observe that $R_{PM}$ will increase with increasing embedding payloads, and $R_{PM}$ is usually less than 67\textpertenthousand \ and 39\textpertenthousand \ for spatial and JPEG steganography separately even the embedding payload is as high as 0.5bpp / 0.50bpnz.

Fig. \ref{fig:violin_plot} shows the violin plots of $R_{PM}$ for HILL and BET-HILL in  different cases.  From the two figures, we observe that the median number of $R_{PM}$ usually increases with increasing payload.  Even when the payload is as high as 0.5 bpp / 0.5 bpnz,  the median number of  $R_{PM}$ is less than 80\textpertenthousand  ~/  40\textpertenthousand, which means that we can achieve great improvement  (refer to Table \ref{tab:security_spatial} and Table \ref{tab:security_jpeg}) via  just modifying a tiny fraction of embedding units for any given stego images $Y$.

\begin{figure*}[!]
  \centering

  \begin{subfigure}{0.33\textwidth}
    \centering
    \includegraphics[width=0.9\linewidth]{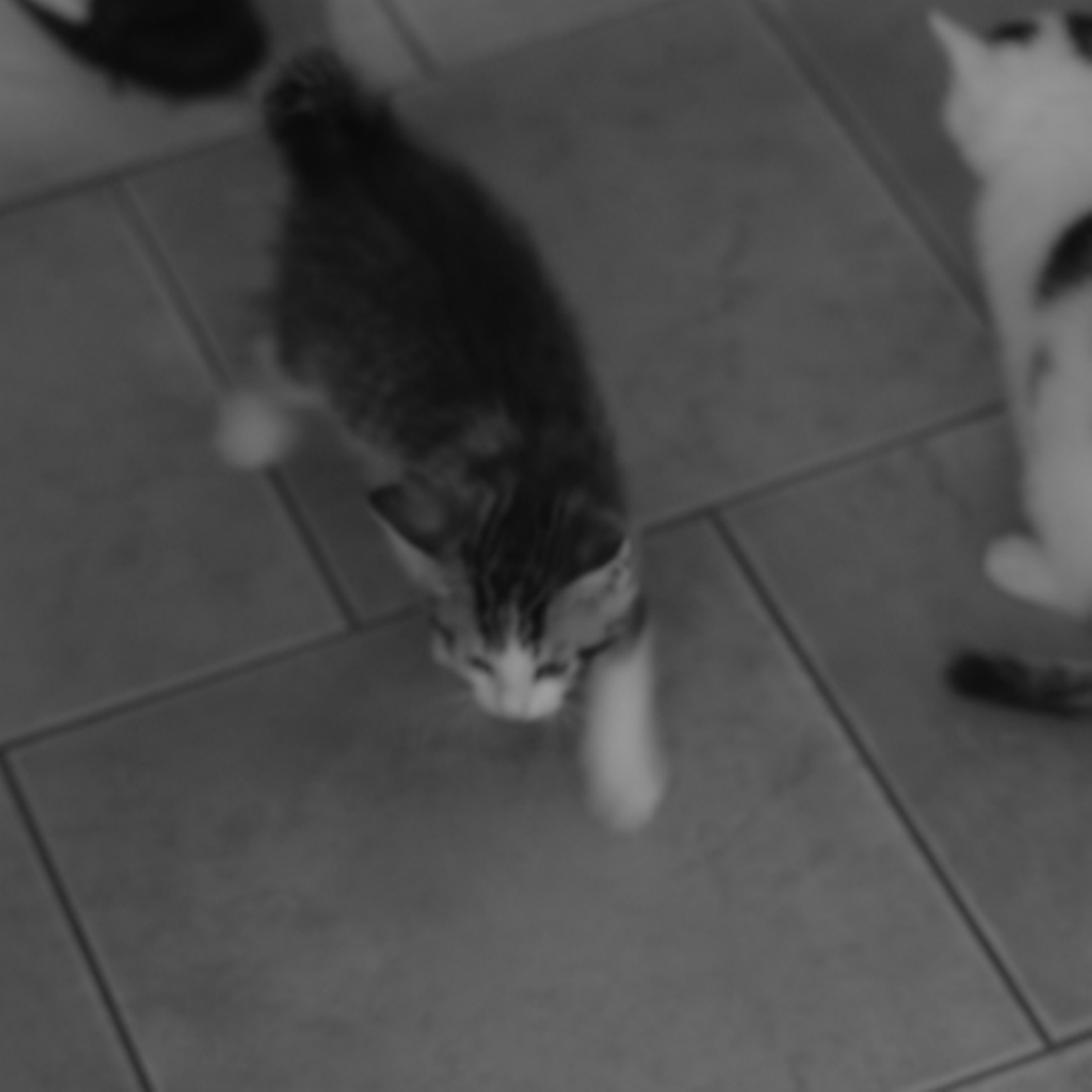}
    \caption{No.7353}
   \end{subfigure}%
  \begin{subfigure}{0.33\textwidth}
    \centering
    \includegraphics[width=0.9\linewidth]{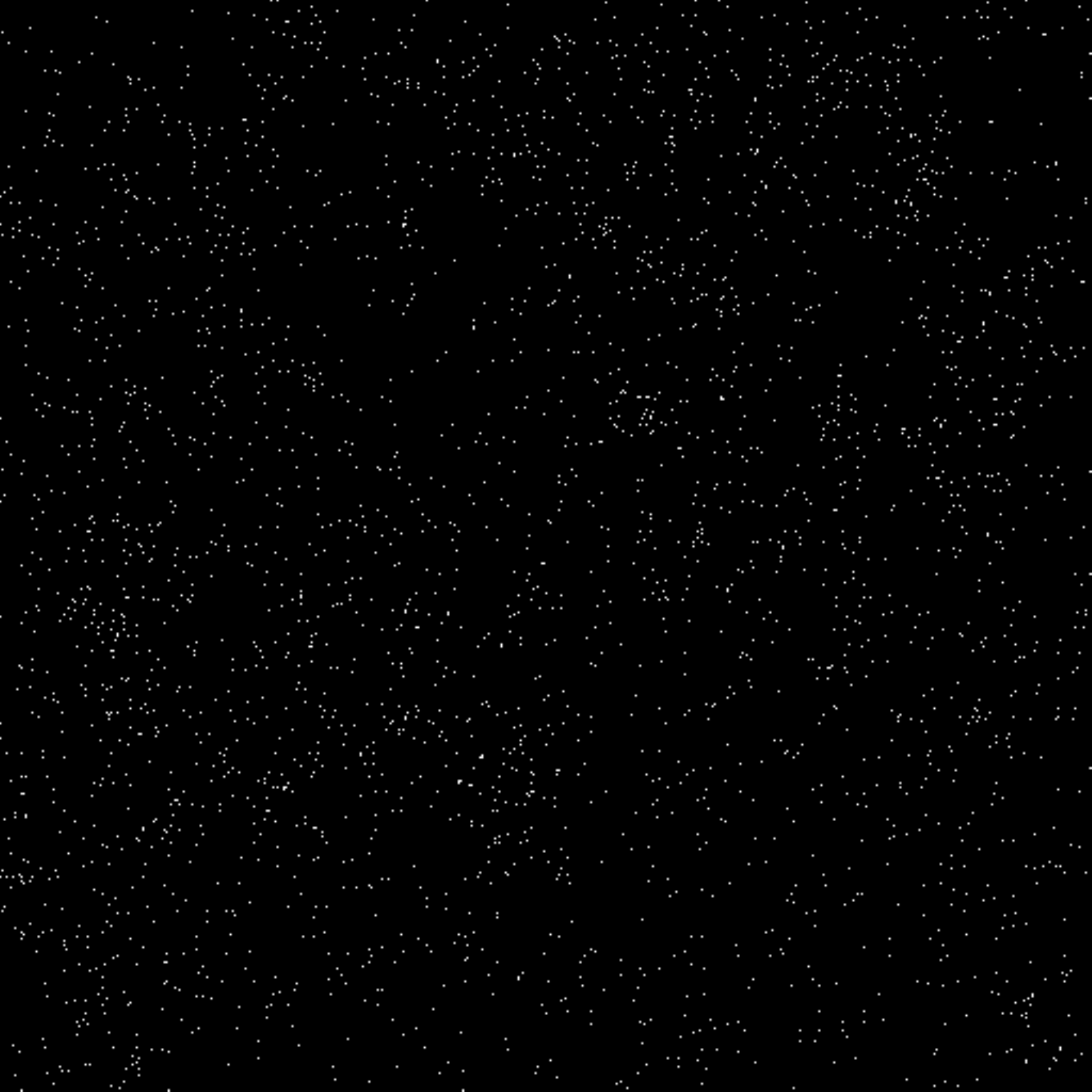}
    \caption{Steganography Modification}
  \end{subfigure}%
  \begin{subfigure}{0.33\textwidth}
    \centering
    \includegraphics[width=0.9\linewidth]{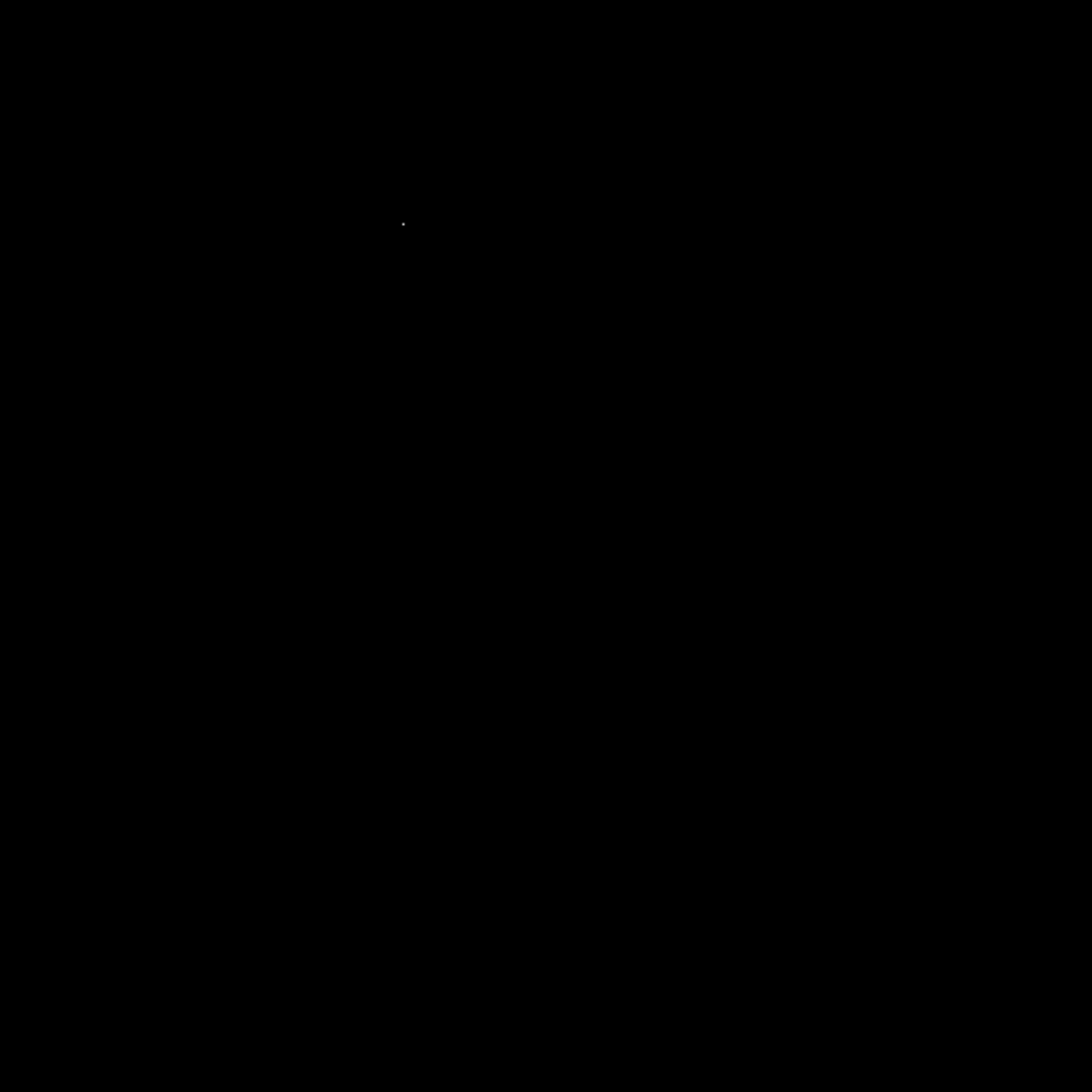}
    \caption{Post-Modification}
  \end{subfigure}
  \begin{subfigure}{0.33\textwidth}
    \centering
    \includegraphics[width=0.9\linewidth]{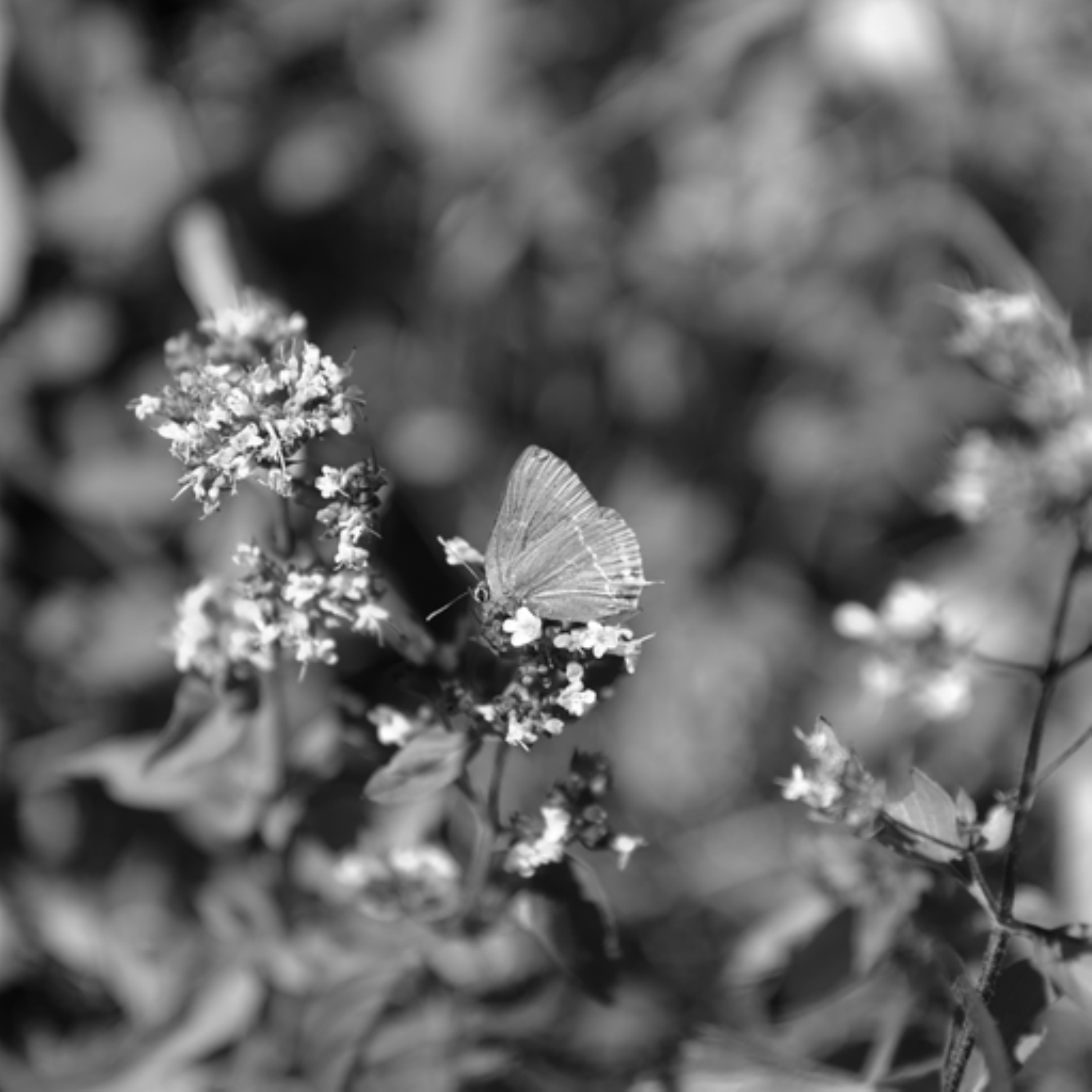}
    \caption{No.1265}
   \end{subfigure}%
  \begin{subfigure}{0.33\textwidth}
    \centering
    \includegraphics[width=0.9\linewidth]{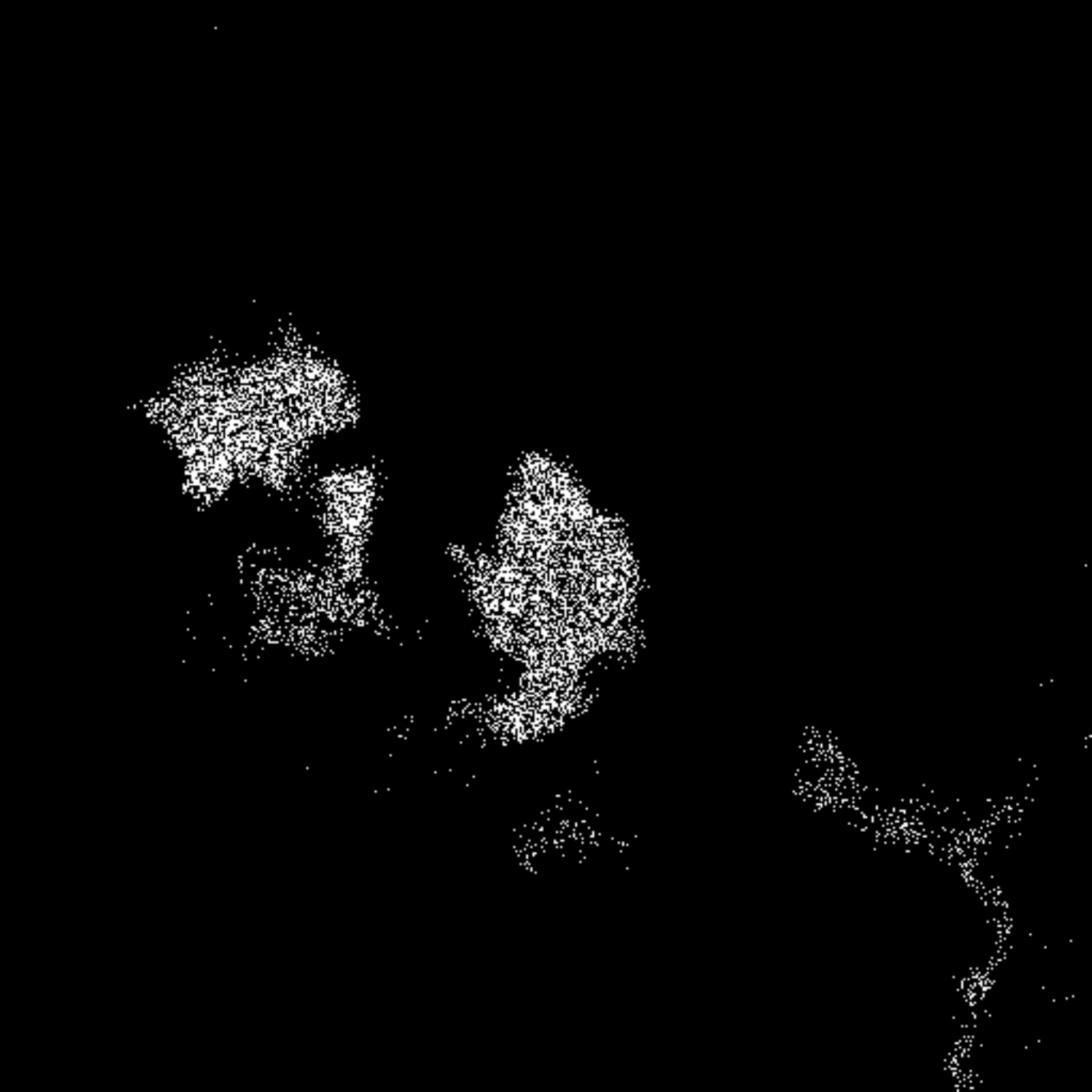}
    \caption{Steganography Modification}
   \end{subfigure}%
  \begin{subfigure}{0.33\textwidth}
    \centering
    \includegraphics[width=0.9\linewidth]{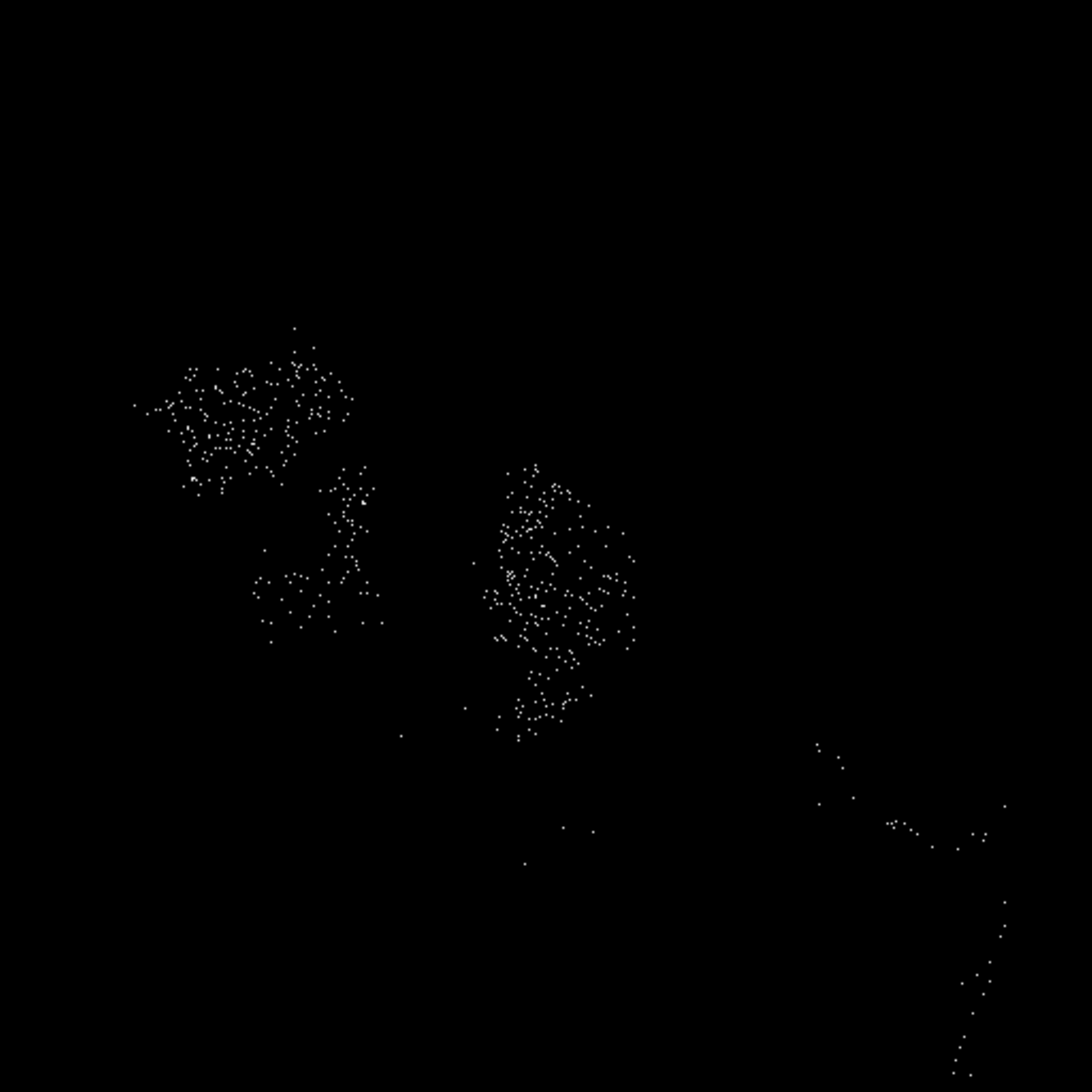}
    \caption{Post-Modification}
  \end{subfigure}%

  \caption{Steganography modification using HILL (0.1 bpp) and the proposed post-modification for two typical image examples. The densities of steganography modification for the two images are 0.05 and 0.27 respectively, and the corresponding numbers of post-modification are 1 and 523 respectively.}
  \label{fig:selected_img_spatial}
\end{figure*}

\vspace{0.5em} \subsubsection{\textbf{Post-Modification Rate  vs.  Density of Steganography Modification}}

\begin{figure*}[!]
  \centering
  \begin{subfigure}{0.5\textwidth}
    \centering
    \includegraphics[width=1\linewidth]{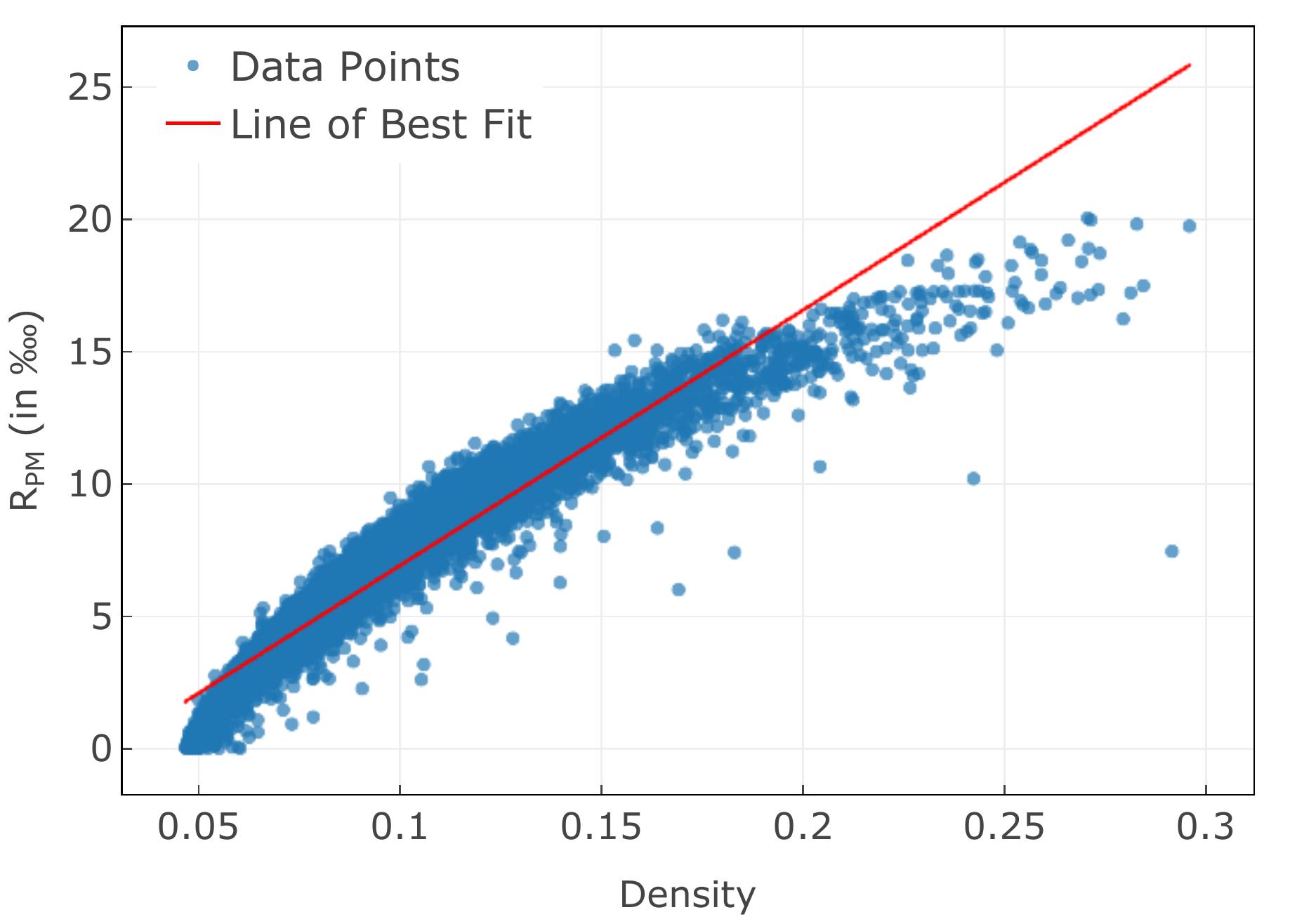}
    \caption{HILL for 0.1 bpp}
   \end{subfigure}%
  \begin{subfigure}{0.5\textwidth}
    \centering
    \includegraphics[width=1\linewidth]{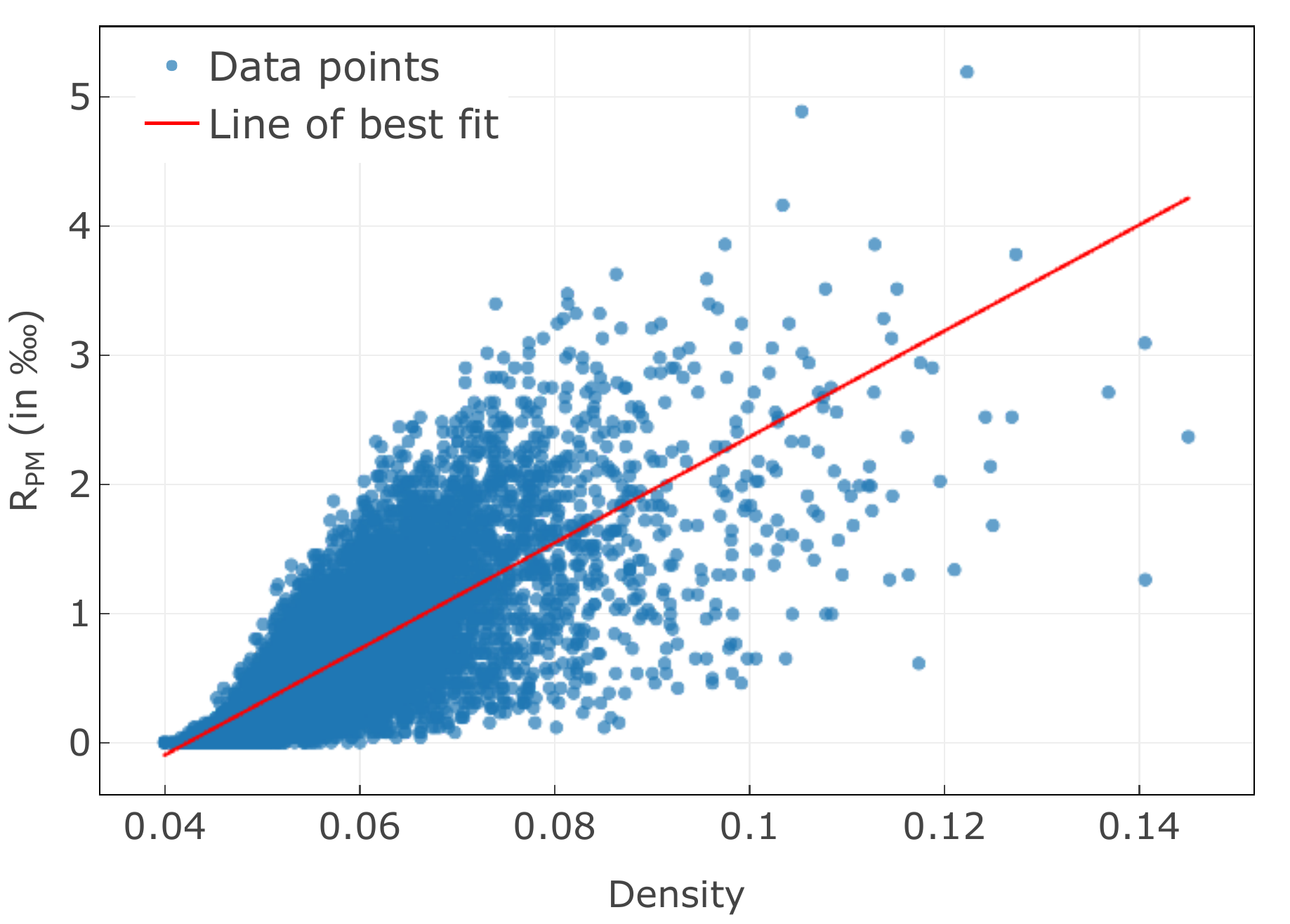}
    \caption{BET-HILL for 0.1 bpnz (QF=75)}
  \end{subfigure}%
  \caption{The scatter plot of  $R_{PM}$ vs. the density of steganography modification, the corresponding Pearson correlation coefficients  for the two cases are  0.95 and 0.70  respectively.}
  \label{fig:scatter_plot}
\end{figure*}

From  Fig. \ref{fig:violin_plot}, we also  observe that for a given payload,  the values of  $R_{PM}$ will change a lot for different images.  
Taking HILL for 0.1 bpp for instance, the minimum of $R_{PM}$ is close to 0, while the maximum become close to  30,  meaning the range of $R_{PM}$ is over 20 in this case.  Furthermore,  the range will increase with increasing payload or quality factor.  In this section, we will analyze the factor which affects the values of $R_{PM}$. 

 Fig. \ref{fig:selected_img_spatial} shows  the  steganography modifications and  the post-modifications of two typical images using HILL for payload 0.1 bpp.   From Fig. \ref{fig:selected_img_spatial}, we  observe that for the first image, the steganography modifications seem uniformly dispersed throughout the whole image,  while it  is highly concentrated on a small part for the second one.  After performing our method, the numbers of the post-modification are 1 and 523 separately.  Thus we expect that there should be a positive correlation between the relative post-modification rate $R_{PM}$ and the density of steganography modification.  To verify this, we define the  density of steganography modification in the following way.  We first compare the difference between cover $X$  and  stego $Y$,   and  divide the difference (i.e. $Y\not=X$) into $5\times 5$ overlapping small blocks.  And then we just consider those blocks which contain steganography modification,  denoted as as $B_i,  i=1, 2 \ldots N$. For each block $B_i$, we  calculate the proportion of steganography modification $|B_i| / 25$,   where $|B_i|$ denotes the number of steganography modification in block $B_i$,  $0< |B_i| / 25 \le 1$.  Finally, we  define the  density of steganography modification for the stego image $Y$  as follows.  
\begin{equation}
\label{eq:d_y}
D_Y = \frac{1}{N} \sum_{n=1}^{N} \frac{|B_i|}{25} 
\end{equation}
Based on this definition,   the densities of  two  images in Fig.\ref{fig:selected_img_spatial} are 0.05 and 0.27 respectively. 
We further calculate  $R_{PM}$ and the  density  for 10,000 images in BOSSBase,  and show the scatter plot for HILL (0.1 bpp) and BET-HILL (0.1 bpnz, QF=75) in Fig \ref{fig:scatter_plot}.  For display purpose, we remove  outlying data (less than 0.15\% with larger values) in this figure.  From Fig. \ref{fig:scatter_plot},  it is obvious that $R_{PM}$ increases with  increasing density.  In this case, the corresponding Pearson correlation coefficients  are  0.95 and 0.70  respectively,  meaning the linear relationships between the $R_{PM}$ and the density  are relatively strong, which  fits our expectation  well.  Similar results can be found for other steganographic methods  and payloads.   

\subsection{Evaluation on Processing Time}

\begin{table}[!t]
  \renewcommand\arraystretch{1.1}
  \caption{Processing time (s) comparison with different  methods in spatial domain.}
  \label{tab:time_spatial}
  \centering
  \begin{tabular}{c||ccccc}
     \hline

     \textbf{} & \textbf{0.1} & \textbf{0.2} & \textbf{0.3} & \textbf{0.4} & \textbf{0.5} \\

    \hline
    \hline
    S-UNI       & 0.30 & 0.28 & 0.29 & 0.28 & 0.29 \\
    S-UNI-SPP          & 0.17 & 0.20 & 0.23 & 0.27 & 0.31 \\
    \hline
    MIPOD        & 1.68 & 1.78 & 1.84 & 1.90 & 1.90 \\
    MIPOD-SPP          & 0.17 & 0.20 & 0.25 & 0.29 & 0.34 \\
    \hline
    HILL        & 0.19 & 0.19 & 0.19 & 0.19 & 0.19 \\
    HILL-SPP          & 0.18 & 0.21 & 0.25 & 0.29 & 0.34 \\
    \hline
    CMD-HILL Embed       & 0.32 & 0.33 & 0.32 & 0.32 & 0.32 \\
    CMD-HILL SPP          & 0.19 & 0.23 & 0.28 & 0.33 & 0.38 \\
    \hline

  \end{tabular}
\end{table}

\begin{table}[!t]
  \renewcommand\arraystretch{1.1}
  \caption{Processing time (s) comparison with different  methods in JPEG domain.}
  \label{tab:time_jpeg}
  \centering
  \begin{tabular}{c||c||ccccc}
     \hline

    \textbf{QF} & \textbf{Steganography} & \textbf{0.1} & \textbf{0.2} & \textbf{0.3} & \textbf{0.4} & \textbf{0.5} \\

    \hline
    \hline

    \multirow{6}{*}{75} & J-UNI        & 3.07 & 3.07 & 3.07 & 3.07 & 3.06 \\
    & J-UNI-SPP          & 0.47 & 0.48 & 0.49 & 0.50 & 0.50 \\
    \cline{2-7}
    & UERD       & 0.08 & 0.08 & 0.07 & 0.07 & 0.07 \\
    & UERD-SPP          & 0.47 & 0.48 & 0.49 & 0.49 & 0.50 \\
    \cline{2-7}
    & BET-HILL        & 0.68 & 0.69 & 0.69 & 0.69 & 0.69 \\
    & BET-HILL-SPP          & 0.47 & 0.48 & 0.48 & 0.49 & 0.50 \\
    \hline

    \multirow{6}{*}{95} & J-UNI Embed       & 3.06 & 3.06 & 3.08 & 3.04 & 3.05 \\
    & J-UNI SPP          & 0.48 & 0.50 & 0.52 & 0.54 & 0.56 \\
    \cline{2-7}
    & UERD         & 0.08 & 0.07 & 0.07 & 0.07 & 0.07 \\
    & UERD SPP          & 0.48 & 0.50 & 0.51 & 0.53 & 0.55 \\
    \cline{2-7}
    & BET-HILL         & 0.69 & 0.74 & 0.71 & 0.71 & 0.71 \\
    & BET-HILL SPP          & 0.48 & 0.50 & 0.51 & 0.53 & 0.56 \\
    \hline

  \end{tabular}
\end{table}

In this section, we will evaluate the processing time of the proposed method. To achieve convincing results, we report the average results on 100 images randomly selected from BOSSBase.  For comparison,  we also provide the processing time of the corresponding steganography method.  The average results are shown in Table \ref{tab:time_spatial} and Table \ref{tab:time_jpeg}.  From the two tables, we have two following observations: 
\begin{itemize}
\item For a given steganography method,   the processing time  usually increases  with  increasing payload since more steganography modification should be dealt with.   Taking HILL for instance,  the time processing  is 0.18s for 0.1 bpp, while it becomes 0.34s for 0.5 bpp. 

\item  For the same reason,  for a given JPEG steganography method and a payload,  the processing time  usually increases  with  increasing quality factor.   Taking  BET-HILL for 0.5 bpnz for instance,  the processing time is 0.50s for QF=75, while it increases to 0.56s for QF=95.

\end{itemize}

Overall, the processing time of the proposed method is very short (less than 0.60s per image in all cases), which is comparable to or even much shorter  than that of the current  steganography method.

\begin{table*}[!]
  \renewcommand\arraystretch{1.1}
  \caption{Detection accuracies improvements (\%) for  spatial steganography on  three image databases (i.e. BOSSBase, BOWS2 and ALASKA).}
  \label{tab:security_spatial_database}
  \centering
  \begin{tabular}{c||c||ccc||ccc||ccc||ccc||c}
     \hline

    \multirow{2}{*}{\textbf{Steganography}} & \multirow{2}{*}{\textbf{Database}} & \multicolumn{3}{c||}{\textbf{SRM}} & \multicolumn{3}{c||}{\textbf{maxSRMd2}}  & \multicolumn{3}{c||}{\textbf{Xu-Net}} & \multicolumn{3}{c||}{\textbf{SRNet}} & \multirow{2}{*}{\textbf{Average}}  \\

    \cline{3-14}

    & & \textbf{0.1} & \textbf{0.3} & \textbf{0.5} & \textbf{0.1} & \textbf{0.3} & \textbf{0.5} & \textbf{0.1} & \textbf{0.3} & \textbf{0.5} & \textbf{0.1} & \textbf{0.3} & \textbf{0.5}\\

    \hline
    \hline
    \multirow{3}{*}{S-UNI} & BOSSBase             & 0.33 & 1.47 & 1.39 & 0.41 & 1.42 & 1.61 & 0.48 & 2.96 & 2.99 & - & - & - & 1.45 \\
    & BOWS2         & 0.23 & 2.33 & 1.97 & 1.66 & 2.73 & 2.11 & 1.00 & 4.66 & 2.98 & - & - & - & 2.19 \\
    & ALASKA         & 0.21 & 0.65 & 1.08 & 0.38 & 1.18 & 0.98 & \textred{\underline{-0.14}} & 2.48 & 3.48 & 0.72 & 0.60 & 0.57 & 1.02 \\
    \hline
    \multirow{3}{*}{MIPOD} & BOSSBase           & \textred{\underline{-0.12}} & 2.33 & 2.68 & 1.41 & 2.77 & 3.01 & 1.08 & 4.31 & 5.20 & - & - & - & 2.52 \\
    & BOWS2       & 0.57 & 3.32 & 4.94 & 1.47 & 4.92 & 4.03 & 1.92 & 5.23 & 5.21 & - & - & - & 3.51 \\
    & ALASKA         & 0.25 & 1.34 & 1.86 & 0.64 & 1.82 & 2.17 & 1.32 & 4.47 & 4.61 & 0.55 & 0.81 & 1.16 & 1.75 \\
    \hline
    \multirow{3}{*}{HILL} & BOSSBase            & 0.56 & 2.76 & 3.20 & 1.61 & 2.24 & 2.69 & 1.75 & 4.48 & 4.73 & - & - & - & 2.67 \\
    & BOWS2        & 1.52 & 4.75 & 5.04 & 2.76 & 4.54 & 4.22 & 2.66 & 5.92 & 5.67 & - & - & - & 4.12 \\
    & ALASKA         & 0.03 & 1.05 & 1.25 & 0.92 & 1.86 & 1.79 & 0.93 & 4.11 & 6.11 & 0.88 & 0.41 & 0.62 & 1.67 \\
    \hline
    \multirow{3}{*}{CMD-HILL} & BOSSBase        & 0.54 & 0.91 & 1.68 & 0.39 & 0.93 & 0.96 & 0.42 & 2.01 & 3.14 & - & - & - & 1.22 \\
    & BOWS2    & 0.65 & 2.71 & 4.12 & 0.94 & 3.00 & 3.67 & 1.96 & 3.22 & 4.64 & - & - & - & 2.77 \\
    & ALASKA         & 0.46 & 0.46 & 1.11 & 0.30 & 1.18 & 1.58 & \textred{\underline{-0.09}} & 1.69 & 2.01 & 0.06 & 0.20 & 0.21 & 0.76 \\
    \hline

  \end{tabular}
\end{table*}

\begin{table*}[!]
  \renewcommand\arraystretch{1.1}
  \caption{Detection accuracies improvements (\%) for  JPEG steganography on  three image databases (i.e. BOSSBase, BOWS2 and ALASKA).}
  \label{tab:security_jpeg_database}
  \centering
  \begin{tabular}{c||c||c||ccc||ccc||ccc||ccc||c}
     \hline

    \multirow{2}{*}{\textbf{QF}} & \multirow{2}{*}{\textbf{Steganography}} & \multirow{2}{*}{\textbf{Database}} & \multicolumn{3}{c||}{\textbf{GFR}} & \multicolumn{3}{c||}{\textbf{SCA-GFR}} & \multicolumn{3}{c||}{\textbf{J-Xu-Net}}  & \multicolumn{3}{c||}{\textbf{SRNet}} & \multirow{2}{*}{\textbf{Average}} \\

    \cline{4-15}

    & & & \textbf{0.1} & \textbf{0.3} & \textbf{0.5} & \textbf{0.1} & \textbf{0.3} & \textbf{0.5} & \textbf{0.1} & \textbf{0.3} & \textbf{0.5} & \textbf{0.1} & \textbf{0.3} & \textbf{0.5} \\

    \hline
    \hline
    \multirow{9}{*}{75} & \multirow{3}{*}{J-UNI} & BOSSBase             & \textred{\underline{-0.03}} & 0.59 & 1.04 & 0.61 & 0.87 & 0.82 & 0.05 & 0.16 & 0.44 & - & - & - & 0.51 \\
    & & BOWS2         & \textred{\underline{-0.16}} & 1.01 & 1.06 & 0.03 & 0.69 & 1.18 & 0.40 & 0.08 & 1.28 & - & - & - & 0.62 \\
    & & ALASKA         & 0.06 & 0.38 & 1.44 & 0.28 & 1.18 & 1.20 & 0.06 & 0.92 & 1.58 & \textred{\underline{-0.24}} & 0.52 & 0.98 & 0.70 \\
    \cline{2-16}
    & \multirow{3}{*}{UERD} & BOSSBase           & 0.82 & 1.16 & 1.60 & 0.29 & 0.87 & 1.05 & 0.38 & 0.43 & \textred{\underline{-0.12}} & - & - & - & 0.72 \\
    & & BOWS2       & \textred{\underline{-0.11}} & 1.07 & 1.43 & 0.63 & 0.96 & 1.08 & 1.25 & 2.35 & 0.52 & - & - & - & 1.02 \\
    & & ALASKA         & 0.47 & 1.08 & 1.65 & 0.14 & 1.21 & 1.77 & 0.22 & 0.82 & 1.80 & 0.13 & 1.01 & 0.27 & 0.88 \\
    \cline{2-16}
    & \multirow{3}{*}{BET-HILL} & BOSSBase            & 0.44 & 1.92 & 1.68 & 1.11 & 2.26 & 1.99 & 1.35 & 1.61 & 1.47 & - & - & - & 1.54 \\
    & & BOWS2        & 0.47 & 1.80 & 2.15 & 1.33 & 2.36 & 1.68 & 0.97 & 2.38 & 1.45 & - & - & - & 1.62 \\
    & & ALASKA         & \textred{\underline{-0.03}} & 1.07 & 1.85 & 0.16 & 1.26 & 2.01 & 0.32 & 1.85 & 3.58 & 0.84 & 2.37 & 2.57 & 1.49 \\
    \hline

    \multirow{9}{*}{95} & \multirow{3}{*}{J-UNI} & BOSSBase             & 0.10 & 0.60 & 1.65 & 0.07 & 1.23 & 1.73 & 0.18 & 1.09 & \textred{\underline{-0.15}} & - & - & - & 0.72 \\
    & & BOWS2         & 0.03 & 0.95 & 2.66 & 0.08 & 1.27 & 2.60 & \textred{\underline{-0.04}} & 1.41 & 1.34 & - & - & - & 1.14 \\
    & & ALASKA         & 0.06 & 0.36 & 1.47 & 0.04 & 0.23 & 1.52 & 0.03 & 0.47 & 1.76 & 0.00 & 0.36 & 1.23 & 0.63 \\
    \cline{2-16}
    & \multirow{3}{*}{UERD} & BOSSBase           & 0.07 & 1.83 & 3.09 & 0.27 & 1.72 & 2.79 & 0.08 & 0.09 & 0.67 & - & - & - & 1.18 \\
    & & BOWS2       & \textred{\underline{-0.11}} & 1.50 & 3.85 & 0.27 & 2.20 & 2.96 & 0.01 & 1.93 & \textred{\underline{-0.17}} & - & - & - & 1.38 \\
    & & ALASKA         & 0.14 & 0.96 & 2.26 & 0.02 & 1.58 & 2.58 & \textred{\underline{-0.15}} & 1.73 & 2.96 & 0.64 & 1.45 & 1.73 & 1.33 \\
    \cline{2-16}
    & \multirow{3}{*}{BET-HILL} & BOSSBase            & 0.18 & 1.08 & 2.73 & 0.72 & 1.97 & 2.77 & 0.57 & 1.27 & 1.93 & - & - & - & 1.47 \\
    & & BOWS2        & \textred{\underline{-0.04}} & 1.08 & 4.13 & 0.23 & 2.09 & 3.05 & 0.20 & 2.32 & 1.64 & - & - & - & 1.63 \\
    & & ALASKA         & 0.19 & 0.89 & 1.81 & 0.24 & 1.03 & 2.17 & \textred{\underline{-0.03}} & 0.42 & 1.91 & 0.02 & 1.68 & 3.21 & 1.13 \\

    \hline

  \end{tabular}
\end{table*}

\subsection{Security Evaluation on Other Image Databases}
In this section,  we will evaluate the security performance on two other databases including 10,000 gray-scale images of size $512 \times 512$ from BOWS2 \cite{bows2} and 80,005 gray-scale images of size $256 \times 256$ \footnote{The images are resampled using "imresize()" in Matlab with default settings from images of size $512\times 512$.} from ALASKA \cite{cogranne2019the}.   For BOWS2,  the partition of image dataset and hyper-parameters are the same as previous ones used for the BOSSBase.  For  ALASKA,  we randomly select 60,005 images for training while the other 20,000 for testing.   In addition,   the current best CNN-based steganalysis,  i.e.,  SRNet \cite{boroumand2018deep},  is used for security evaluation on ALASKA\footnote{Since SRNet is originally designed for images of size $256\times 256$,  and it needs sufficient training data to get good results, we do not use SRNet to evaluate the security on BOSSBase and BOWS2. }.  For comparison, the detection accuracy improvements on the three image databases (i.e. BOSSBase, BOWS2 and ALASKA) are shown in Table \ref{tab:security_spatial_database} and Table \ref{tab:security_jpeg_database}. From the two tables, we obtain two following observations:

\begin{itemize}
\item  Almost in all cases, the proposed method can effectively enhance the steganography security.  In many  cases, we can achieve over 3\% and 2\% for spatial and JPEG domains separately, which is a significant improvement on modern steganography.  In a few cases, the  security performance will drop slightly (less than 0.24\%).  On average, our method can enhance the steganography security for all cases (refer to the final column in two tables) .  

\item  The improvement is different for the three image datasets.  Compared to the results on BOSSBase and ALASKA, we can achieve  greater improvements on  BOWS2 in many cases,  especially for steganography methods in spatial domain.  In addition, the improvement would changes  for different steganalytic methods.  The improvement evaluated on the current best CNN-based steganalytic detector (i.e., SRNet) is relatively smaller than that on the there other detectors. 
\end{itemize}

 \begin{table}[!]
  \renewcommand\arraystretch{1.1}
  \caption{Detection accuracies improvements (\%) of our previous method   \cite{chen2019enhancing}  and  the proposed method for different steganography methods in spatial domain (0.4 bpp). }
  \label{tab:ih_spatial}
  \centering
  \begin{tabular}{c||cccc|c}
     \hline
    \textbf{Steganography} & \textbf{S-UNI} & \textbf{MIPOD} & \textbf{HILL} & \textbf{CMD-HILL} & \textbf{Average} \\
    \hline
    \hline
    Method \cite{chen2019enhancing}  & 0.96 & 0.85 & 1.69 & 0.10 & 0.90 \\ 
    Proposed  & \textblue{1.37*} & \textblue{2.47*} & \textblue{3.30*} & \textblue{1.13*} & \textblue{2.07*} \\ 
    \hline
  \end{tabular}
\end{table}

\begin{table}[!]
  \renewcommand\arraystretch{1.1}
  \caption{Detection accuracies improvements (\%) of our previous method   \cite{chen2019enhancing}  and  the proposed method for different steganography methods in JPEG domain (0.4 bpnz QF=95) .}
  \label{tab:ih_jpeg}
  \centering
  \begin{tabular}{c||ccc|c}
     \hline
    \textbf{Steganography} & \textbf{J-UNI} & \textbf{UERD} & \textbf{BET-HILL} & \textbf{Average} \\
    \hline
    \hline
    Method \cite{chen2019enhancing}        & \textred{\underline{-0.19}} & \textred{\underline{-0.11}} & 0.06 & \textred{\underline{-0.08}} \\ 
    Proposed  & \textblue{1.04*} & \textblue{2.54*} & \textblue{2.34*} & \textblue{1.97*} \\ 
    \hline
  \end{tabular}
\end{table}

\subsection{Comparison with Our Previous Method}

In this section, we will compare the steganography security and the processing time with our previous work 
 \cite{chen2019enhancing}.  
 
\vspace{0.5em} \subsubsection{\textbf{Comparison on Security Performance}}  For simplification,  we evaluate  spatial steganography methods for payload 0.4 bpp using SRM, and evaluate JPEG steganography methods for payload 0.4 bpnz with QF 95 using GFR both on BOSSBase.  The experimental results are shown in Table \ref{tab:ih_spatial} and Table \ref{tab:ih_jpeg} separately.  From the two tables, we observe that in spatial domain, both methods can enhance steganography security.  On average,  the previous method and the proposed method  achieve an improvement of about 1\% and 2\% separately.  In JPEG domain,  our previous method  \cite{chen2019enhancing} does not work effectively,  while the proposed method  still  achieves an average improvement of about 2\%.   

\vspace{0.5em} \subsubsection{\textbf{Comparison on Processing Time}}  The average processing time  evaluated on 100 randomly selected images from BOSSBase  are shown in Table \ref{tab:ih_spatial_time} and  Table \ref{tab:ih_jpeg_time} separately.
From the two tables,  we  observe that the proposed method is significantly faster than the previous method.  On average, the proposed method are able to achieve about 7 times acceleration in spatial domain, and about 5 times acceleration in JPEG domain. 

The above results show that the proposed method is much more effective and faster than the previous one  \cite{chen2019enhancing} \footnote{Note that the fast method for updating image residual in section \ref{subsubsec:Updating} is also employed in our previous work for comparison. }.

\begin{table}[!]
  \renewcommand\arraystretch{1.1}
  \caption{Processing time (s) for  our previous method   \cite{chen2019enhancing}  and  the proposed method for  spatial steganography (0.4 bpp).  }
  \label{tab:ih_spatial_time}
  \centering
  \begin{tabular}{c||cccc|c}
     \hline
    \textbf{Steganography} & \textbf{S-UNI} & \textbf{MIPOD} & \textbf{HILL} & \textbf{CMD-HILL} & \textbf{Average} \\
    \hline
    \hline
    Method  \cite{chen2019enhancing}      & 2.03 & 2.03 & 2.03 & 2.03 & 2.03 \\ 
    Proposed  & 0.27 & 0.29 & 0.29 & 0.33 & 0.30 \\ 
    \hline
  \end{tabular}
\end{table}

\begin{table}[!]
  \renewcommand\arraystretch{1.1}
  \caption{Processing time (s) for   our previous method   \cite{chen2019enhancing}  and  the proposed method for  JPEG steganography (0.4 bpnz QF=95)  }
  \label{tab:ih_jpeg_time}
  \centering
  \begin{tabular}{c||ccc|c}
     \hline
    \textbf{Steganography} & \textbf{J-UNI} & \textbf{UERD} & \textbf{BET-HILL} & \textbf{Average} \\
    \hline
    \hline
    Method  \cite{chen2019enhancing}      & 2.79 & 2.78 & 2.78 & 2.78 \\ 
    Proposed  & 0.54 & 0.53 & 0.53 & 0.53 \\ 
    \hline
  \end{tabular}
\end{table}

\section{Conclusion}
\label{sec:conclusion}
In this paper, we propose a novel method to  enhance the steganography security via  stego post-processing. The main contributions of this paper are as follows.

\begin{itemize}
\item Unlike existing works which focus on  embedding costs design (e.g.,  HILL and UNIWARD) or enhancement  (e.g., CMD and BBC) according to some predetermined rules (such as complexity-first, spreading, clustering \cite{li2014investigation},  and preserving the continuity of  block boundary) during data embedding,  the  proposed method  tries to directly modify embedding units of stego to reduce the residual distance when the embedding processing is completed with an existing steganography.  
\vspace{0.5em}
\item The proposed method is universal,  because it can be effectively applied in  those steganographic methods using STCs for data hiding,  including most modern image steganography methods both in spatial and JPEG domains.  In addition, the number of modified  embedding units is tiny with the proposed method. 
\vspace{0.5em}
\item  On average, the proposed method  can enhance the steganography security  in all cases.  In many cases, we can even achieve over  3\% and 2\%  for the modern steganographic methods  in  spatial and JPEG domain separately.  Note that such an improvement is significant, especially for enhancing the advanced steganography, such as CMD-HILL and BET-HILL.  
\end{itemize}

In our experiments, we try to reduce the Manhattan distance between cover residual  and stego resudial via post-modification.  Other steganalytic measures,  such as the co-occurrence matrices of image residual in SRM  and some deep learning based features will be considered in our future work.  In addition,  we will combine the technique of  adversarial example to further improve the steganography security.

\balance
\bibliographystyle{IEEEtran}
\bibliography{uni_spp}
\end{document}